\documentclass[10pt]{IEEEtran}
\usepackage{bm,upgreek}
\usepackage{graphicx}
\graphicspath{{Figures/}}
\usepackage{amsmath}
\usepackage{amssymb}
\usepackage[usenames,dvipsnames]{xcolor}
\usepackage{xspace}
\usepackage{multirow}
\usepackage{cite}
\include{localstyle}
\usepackage[normalem]{ulem}
\usepackage{mathrsfs}
\usepackage{amsbsy}
\usepackage{subfigure}

\newcommand{\cg}{\color{red}}
\newcommand{\cgb}{\color{blue}}

\renewcommand{\cg}{}
\renewcommand{\cgb}{}

\newcommand{\hu}{\color{blue}}
\newcommand{\zt}{\color{brown}}
\renewcommand{\hu}{}
\renewcommand{\zt}{}
\newcommand{\mssat}{\texttt{AC-SAT}\xspace}
\newcommand{\av}{\texttt{AVC}\xspace}
\newcommand{\dv}{\texttt{DVC}\xspace}
\newcommand{\sd}{\texttt{SDC}\xspace}
\newcommand{\cn}{{\mbox{$\mathscr{N}$}}\xspace}
\newcommand{\cm}{{\mbox{$\mathscr{M}$}}\xspace}

\begin{document}

\title{Efficient Analog Circuits for Boolean Satisfiability}
\author{Xunzhao Yin,~\IEEEmembership{Student,~IEEE,} 
				Behnam Sedighi, 
				Melinda Varga, \\
				M\'aria Ercsey-Ravasz,
        Zolt\'an Toroczkai,
				Xiaobo Sharon Hu,~\IEEEmembership{Fellow,~IEEE,}
  \thanks{X. Yin and X.S. Hu are with the
    Department of Computer Science and Engineering, University of
   Notre Dame, IN 46556, USA (xyin1, shu@nd.edu).}% <-this % stops a space
  \thanks{B. Sedighi is with Qualcomm, San Diego, CA 92121, USA (behnam.sedighi@gmail.com).}% <-this % stops a space
  \thanks{M. Varga and Z. Toroczkai are with the Department of Physics and the Interdisciplinary 
Center for Network Science and  Applications (iCeNSA),   
University of Notre Dame, Notre Dame, IN, 46556 USA (mvarga, toro@nd.edu).}% <-this % stops a space
  \thanks{M. Ercsey-Ravasz is with the Faculty of Physics, Hungarian Physics Institute, 
Babes-Bolyai University,  Cluj-Napoca, Romania, (ercsey.ravasz@phys.ubbcluj.ro)}
\thanks{Color versions of one or more of the figures in this paper are available online at http://ieeexplore.ieee.org.}
\thanks{Digital Object identifier 10.1109/TVLSI.2017.2754192}
}
\IEEEpubid{\begin{minipage}{\textwidth}\ \\[12pt]  \centering
	1063-8210 \copyright~ 2017 IEEE. Personal use is permitted, but republication/redistribution requires IEEE permission.\\
	See http://www.ieee.org/publications\_standards/publications/rights/index.html for more information.\end{minipage}}
\maketitle

%\vspace*{-1in}
\begin{abstract}

Efficient solutions to NP-complete problems would significantly benefit both science and industry. %, even affecting aspects of our daily life, from flight scheduling to cyber security. 
However, such problems are intractable on digital computers based on the von Neumann architecture, thus creating the need for alternative solutions to tackle such problems. Recently, a deterministic, continuous-time dynamical system (CTDS) was proposed~\cite{NatPhys_ET11} to solve a representative NP-complete problem, Boolean Satisfiability (SAT). This solver shows polynomial analog time-complexity on even the hardest benchmark $k$-SAT ($k \geq 3$) formulas, but at an energy cost through exponentially driven auxiliary variables. This paper presents a novel analog hardware SAT solver, \mssat,  implementing the CTDS {\cg via incorporating novel, analog circuit design ideas. \mssat} is intended to be used as a co-processor and is programmable for handling different problem specifications. {\cgb It is especially effective for solving hard $k$-SAT problem instances that are challenging for algorithms running on digital machines.}
  %solve any 3-SAT problem with the numbers of variables and constraints equal to or less than that can be handled by a given \mssat implementation. 
Furthermore, with its  modular design, \mssat can {\cg readily be} extended to solve larger size problems, {\cgb while the size of the circuit grows linearly with the product of the number of variables and number of clauses}. The circuit is designed and simulated based on a 32nm CMOS technology. SPICE simulation results show speedup factors of $\sim$10$^4$ on even the hardest 3-SAT problems, when compared with a state-of-the-art SAT solver on digital computers. As an example, for hard problems with $N=50$ variables and $M=212$ clauses, solutions are found within from a few $ns$ to a few hundred $ns$.% with an average power consumption of ${130}$ $mW$.
% the latter due to mainly static power dissipation by analog inverters. 
\end{abstract}

%\input{introduction}
% !TEX root = main.tex
\section{Introduction}
\label{sec:introduction}

With Moore's Law coming to end~\cite{Nature_Waldrop}, exploring novel computational paradigms (e.g., quantum computing and neuromorphic computing) is more imperative than ever. While quantum computing is a promising venue, it is far from being brought to practical reality, with many challenges still to be faced, both in physics and engineering. Neuromorphic computing systems, e.g., Cellular Neural Networks (CNNs)~\cite{IEEETCS_CRV93, IEEETCS_CY88_1, lou2015tfet} and IBM's TrueNorth~\cite{merolla2014truenorth}, have been shown to be promising alternatives for solving a range of problems in, say, sensory processing (vision, pattern recognition) and robotics. Analog {\cg mixed-signal} information processing systems such as CNNs can offer extremely power/energy efficient solutions to some problems that are costly to solve by digital computers~\cite{siegelmann2012neural}. Such systems have received increasing attention in recent years (e.g., ~\cite{daniel2013synthetic, lu2015learning, st2014general}), including parallel analog implementations, see \cite{NatCom_MMI15}.

\IEEEpubidadjcol

In analog computing~\cite{LKIDD_aVLSI02}, the algorithm (representing the ``software'') is a dynamical system often 
expressed in the form of differential equations running in continuous time over real numbers, and its physical implementation 
(the ``hardware'') is any physical system, such as an analog circuit, whose behavior is described by the corresponding dynamical system. 
The equations of the dynamical system are designed such that the solutions to problems appear as attractors for the dynamics 
and the output of the computation is the set of convergent states to those attractors~\cite{JCompl_BSF02}. Although it has been 
shown that systems of ordinary differential equations can simulate any Turing machine~\cite{IEEEWPC_B94, IEEETCS_CRV93, Science_S95}, 
and hence they are computationally universal, they have not yet gained widespread popularity due to the fact that  designing such 
systems is problem specific and usually difficult.
% to design.
However, if an efficient analog engine can be designed to solve NP-complete problems, then according to  the Cook-Levin theorem~\cite{GareyJohnson79}, it would help solve efficiently {\em all} problems in the NP class, as well as benefit a very large number of applications, both in science and engineering.  
%Assuming the strongly held belief that P$\neq$NP, NP-complete problems are intractable on digital machines, meaning that there are no algorithms
%running on a sequential Turing machine that would find a solution in time that scales polynomially with input size (e.g., in the number of variables $N$). 

In this paper, we consider designing analog circuits for solving a representative NP-complete problem, the Boolean satisfiability (SAT) problem.
SAT is quintessential to many electronic design automation problems, and is also at the heart of many decision, scheduling, error-correction and security applications. Here we focus on $k$-SAT, 
%in which we need to find an assignment to $N$
%Boolean variables $x_i \in \{0,1\}$, $i=1,\ldots,N$, such that they satisfy a given propositional formula $F$. This formula in conjunctive normal form (CNF) 
%is expressed as the conjunction of $M$ clauses $C_m$, $m=1,\ldots,M$, i.e., $F = \bigwedge_{m=1}^{M} C_m$, where each clause is formed by the disjunction of $k$ literals (which are variables or their complements).  An example of a clause for 3-SAT would be $C_5 = (x_3\vee \overline{x}_{19} \vee x_{53})$. 
for which is well known {\cg to be NP-complete for $k \geq 3$~\cite{GareyJohnson79}}. 
%The hardness comes from the e xponential size of the search space: the simplest Turing-machine algorithm, which sequentially searches all the assignments, and thus it does not exploit the structure of the search space,  clearly has a worst-case complexity of $2^N$. 
The currently best known  deterministic, sequential discrete algorithm that exploits some properties of the search space has a worst case complexity of ${\cal O}(1.473^N)$~\cite{TheorCompSci_BK04}. Other algorithms are based on heuristics and while they may perform well on some SAT formula classes, there are always other formulas on which they take exponentially long times or get stuck indefinitely.
Some of the better known SAT solvers include Zchaff~\cite{moskewicz2001chaff}, MiniSat~\cite{een2003extensible},  RSat~\cite{pipatsrisawat2007rsat}, WalkSAT~\cite{Selman96localsearch}, Focused Record-to-Record Travel (FRRT)~\cite{Dueck93newopt} and Focused Metropolis Search (FMS)~\cite{JStatMech_SeitzAO_2005}. They typically consist of decision, deduction, conflict analysis and other functions~\cite{davis1962machine} that employ the capability of digital computers to assign values to literals, conduct  Boolean constraint propagation (BCP) and backtrack  conflicts~\cite{silva1997grasp, zhang2001efficient}.  

A number of hardware based SAT solvers have been proposed in the past. FPGAs based solutions have been investigated to accelerate the BCP part found in all ``chaff-like'' modern
SAT solvers~\cite{davis2008designing, davis2008practical,  thong2013fpga}. Speedups of anywhere between 3X and 38X have been reported when comparing these FPGA based solvers over MiniSat~\cite{een2003extensible}, a well known, high-performance software solver. %previous attempts~\cite{suyama2001solving, zhong1999using, zhong1998solving} imply that it is impossible to completely map a software SAT solver to pure hardware because these solvers are originally proposed at software level, without taking any advantages of hardware level into consideration at all. Fixed solving steps without any deviations in theory side, and limited improvement space for each step in software and hardware efforts together result in only constant development of SAT solvers.
A  custom digital integrated circuit (IC) based SAT solver,  which implements a variant of general responsibility assignment software patterns (GRASP) and accelerates traversal of the implication graph and conflict clause generation, has been introduced in~\cite{gulati2010accelerating, gulati2008efficient}.  A speed up of $\sim$10$^3$ over MiniSat was reported based on simulation together with extrapolation.
Performance of these hardware based approaches still have a lot of room for improvement since the algorithms that these hardware accelerators  are based on are designed for digital computers and thus can typically expect to achieve limited speedup.

Recently, an analog SAT solver circuit was introduced in ~\cite{smithanalog} using the theoretical proposal from \cite{molnar2013asymmetric} based on the CNN architecture. However, the theory in \cite{molnar2013asymmetric} has exponential analog-time complexity, and thus is much less efficient than the  SAT solver 
from \cite{NatPhys_ET11}, which forms the basis for this paper. Furthermore, the circuit from \cite{smithanalog} seems to have been implemented only for a $4 \times 4$ problem size, and no hardware simulation and comparison results were reported.

Ref.~\cite{NatCom_MMI15} proposes a distributed mixed (analog and digital) algorithm that is implementable on VLSI devices. It is based on a heuristic method combined with stochastic  search, drawing on the natural incommensurability of analog oscillators. Assuming P$\ne$NP, in order to have efficient, polynomially scaling solution times, one would require exponentially many computing elements, that is, exponentially scaling hardware resources. However,  the method in Ref. \cite{NatPhys_ET11} trades time-cost for \textit{energy-cost}, which in practical  terms is preferable to massive amounts of hardware resources. It is quite possible that from an engineering point of view the ideal approach combines both types of tradeoffs: time vs energy and time vs hardware (distributed). The heuristic stochastic search in \cite{NatCom_MMI15} is effectively a simulated annealing method, which implies high exponential runtimes for worst case formulas. In contrast, the analog approach in \cite{NatPhys_ET11} is fully deterministic and extracts maximum information about the solution, embedded implicitly within the system of clauses and can solve efficiently the hardest benchmark SAT problems - at an energetic cost \cite{NatPhys_ET11}.

%Note that hardness is a statement about computational resources needed to find solutions, not only about running time: naively, we may consider $2^N$ computers, each testing a different assignment, and thus solutions will be found in exactly one step. In general, NP-complete problems are expected to require (in the worst-case) exponentially scaling resources in time and/or energy. However, since we can't control/generate time, but we may generate energy (within limits), this opens up the possibility for smart physical systems that push the boundaries of practical computability by balancing the costs of computation between the physical time and the energy variables {\em within the same device}. The formulation of the ultimate time-energy boundary for computability by physical devices is an open question, however, the current work is an indication that there is still significant room for improvement before this boundary is reached.  

Here we propose a novel analog   {\em hardware} SAT solver, referred to as \mssat\footnote{We refer to \mssat as an Analog Circuit SAT solver since its main processing engine is analog. However, the entire system is a mixed-signal one as a digital verification component is also included in the hardware system.}. %, is inspired by a recently developed deterministic SAT solver theory. 
\mssat is based on  the deterministic, continuous-time dynamical system (CTDS) in the form of coupled ordinary differential equations  presented in \cite{NatPhys_ET11}.  As mentioned above, this system finds SAT solutions in analog polynomial time, however, at the expense of auxiliary variables that 
can grow exponentially, when needed (see Refs \cite{NatPhys_ET11},  \cite{SciRep_ET12} for details).
Though this CTDS is an incomplete solver, it does minimize the number of unsatisfied clauses when there are no solutions, and thus it is also a MaxSAT solver.
The overall design of  \mssat is {\em programmable and modular}, thus it can readily solve any SAT problem of size equal or less than  what is imposed by the hardware limitations, and can also be easily extended to solve larger problems.  
Moreover, to avoid resource-costly implementations of the complex differential equations in CTDS, we introduce a number of novel, analog circuit implementation ideas which lead to much smaller amount of hardware than straightforward implementations, while preserving the critical deterministic behavioral properties of CTDS equations.

We have  validated our design through SPICE  simulations. Our {\zt simulations} show that \mssat can significantly outperform (over tens of thousands times faster than) MiniSat, with the latter running on the latest, high-performance digital processors.  For hard SAT problems with 50 variables and over 200 clauses, compared with the projected performance of a possible custom hardware implementation based a recent FPGA  solver~\cite{thong2013fpga}, \mssat offers more than  $\sim$600X speedup.  {\hu Monte Carlo simulations further demonstrate that \mssat  is robust against device variations.} 

In the rest of the paper, we first review the basic CTDS theory and some of its variants in Section~\ref{sec:background}.  Section~\ref{sec:framework} introduces  the overall \mssat design. In Sec.~\ref{sec:improvedesign},  we  present two alternative designs for a specific component in \mssat .
%Sec.~\ref{sec:evaluation} first shows validation results by comparing \mssat with a software implementation of the CTDS SAT solver, and then summarizes performance, complexity and  capability results for \mssat and compares it with  another software solver MiniSat. 
Sec.~\ref{sec:evaluation} first discusses  simulation-based validation of \mssat and compares the different component designs,  and then summarizes performance results for \mssat with respect to a software implementation of the CTDS SAT solver and MiniSat.
Finally, we conclude the paper in Section~\ref{sec:discussions}.

%\input{background}
% !TEX root = main.tex
\section{Background}
\label{sec:background}
Solving a $k$-SAT problem is to find an assignment to $N$ Boolean variables $x_i \in \{0,1\}$, $i=1,\ldots,N$, such that they satisfy a given propositional formula $F$. $F$ in conjunctive normal form (CNF) is expressed as the conjunction of $M$ clauses $C_m$, $m=1,\ldots,M$, i.e., $F = \bigwedge_{m=1}^{M} C_m$, where each clause is formed by the disjunction of $k$ literals (which are variables or their complements).  An example of a clause for 3-SAT would be $C_5 = (x_3\vee \overline{x}_{19} \vee x_{53})$. 
Following \cite{NatPhys_ET11},   an analog variable 
$s_i$, which can take any real value in the range {$s_i \in [-1,1]$,} is associated  with the Boolean variable $x_i$ such that $s_i = -1$ corresponds to
$x_i$ being FALSE ($x_i = 0$) and $s_i = 1$ to $x_i$ being TRUE
($x_i = 1$). The formula $F = \bigwedge_{m=1}^{M} C_m$ can be encoded via the $M\times N$ matrix 
$\mathbf{C} = \{c_{m,i}\}$ with $c_{m,i} = 1$ when $x_i$ appears in clause $C_m$, $c_{m,i} = -1$ when its complement  $\overline{x}_i$ (negation of $x_i$) appears in $C_m$ and $c_{m,i} = 0$ when
neither appears in $C_m$.
To every clause $C_m$, we associate an analog function $K_m(\mathbf{s}) \in [0,1]$ given by
\begin{equation}
K_{m}(\mathbf{s}) = 2^{-k}\prod_{i=1}^{N} (1-c_{m,i}s_{i})\;. \label{Km}
\end{equation}
It is easy to see that clause $C_m$ is satisfied, iff $K_{m} = 0$. 
Defining a ``potential energy'' function
\begin{equation}
V(\mathbf{s},\mathbf{a}) = \sum_{m=1}^{M} a_{m} K_{m}^2\;, \label{pot}
\end{equation}
where $a_{m} > 0$ are auxiliary variables, {\cg one can see} that all the clauses are satisfied iff $V = 0$. Thus the SAT problem can be reformulated as search in $\mathbf{s}$ for the global minima of $V$ (since the condition $V \geq 0$ always applies). If the auxiliary variables $a_m$ are kept as constants, then for  most hard problems any hill-descending deterministic algorithm (which evolves the variables $s_i(t)$) would eventually become stuck in local minima of $V$ and not find solutions. To avoid this,  the auxiliary variables are endowed with a time-dependence coupled to the analog clause functions $K_m$. Ref  \cite{NatPhys_ET11} proposed 
\begin{eqnarray} 
&&\dot{s}_i = \frac{ds_{i}}{dt} = -\frac{\partial }{\partial s_{i}} 
V(\mathbf{s},\mathbf{a})\;,\;\; i=1,\ldots,N  \label{dyn_s} \\
&&\dot{a}_m = \frac{da_{m}}{dt} = a_{m}K_{m}\;,\;\;\;\;\;\; m=1,\ldots,M \label{dyn_a}
\end{eqnarray}
in which~\eqref{dyn_s} describes a gradient descent on $V$ and \eqref{dyn_a} is an exponential growth driven by the level of non-satisfiability in $K_m$ (which also guarantees that $a_m(t) > 0$, at all times). 
~(\ref{dyn_s}) can be rewritten as
\begin{equation}
\label{dyn_s1}
\frac{ds_{i}}{dt} = \sum_{m=1}^{M} a_{m} D_{m,i}
\end{equation}
where 
\begin{multline}
\label{Dm1}
D_{m,i} = -\frac{\partial }{\partial s_{i}} K_{m}^2= 2K_{m} c_{m,i} \prod_{\substack{
		j=1 \\
		j\neq i}
}^{N} (1-c_{m,j}s_{j}) \;.\\
\end{multline}

\noindent
For the auxiliary variables $a_m$, the formal solution to \eqref{dyn_a} is 
\begin{equation}
a_m(t) = a_m(0) e^{\int_{0}^t d\tau K_m(\mathbf{s}(\tau))} \;, \label{aexp}
\end{equation}
and thus the expression \eqref{pot} of $V$ is dominated by those $K_m$ terms which have been unsatisfied for the longest time during the dynamics, resulting in an analog version of a focused search-type \cite{JStatMech_SeitzAO_2005} dynamics. Also note that system \eqref{dyn_s} - \eqref{dyn_a} is not unique, however, it is simple from a theoretical point of view, and incorporates the necessary ingredients for solving arbitrary SAT problems, due to the exponentially accelerated auxiliary variables.  For details on the performance of the algorithm see  \cite{NatPhys_ET11}. 

It is important to observe that while the scaling of the analog time $t$ to find solutions is polynomial, in hardware implementations, the $a_m$ variables represent voltages or currents and thus the energetic resources needed to find solutions may become exponential for hard formulas which is, of course necessary, assuming P$\neq$NP. However, the $a_m$ variables do not need to grow exponentially all the time and unlimitedly, as in \eqref{dyn_a} and for that reason form \eqref{dyn_a} is not ideal for physical implementations. The challenge is then finding other variants that still significantly outperform digital algorithms, yet they are feasible in terms of physical implementations and costs. Note that such systems as ours essentially convert time costs into energy costs.

Here we introduce another form with the help of time-delays, which, however, {\cg still} keeps the focused nature of the search dynamics but allows the $a_m$'s to \textit{decrease} as well when the corresponding clauses are (nearly) satisfied:
\begin{equation}
\label{dyn_a_new}
\frac{da_{m}}{dt} = a_{m}\{K_{m}(\textbf{s}(t))-[1-\dot{\delta}_m(t)]K_m(\textbf{s}(t-\delta_m(t)))\},
\forall m
\end{equation}
with
\begin{equation}
\label{newaminitial}
a_m(0)>0, \mbox{ and }   \delta_m(0)=0, \forall m
\end{equation}
where the delay functions $\delta_m(t)\in [0,t]$ determine the history window of $K_m(\textbf{s}(t))$  trajectory that has impact on the variation of $a_m$. The formal solution to \eqref{dyn_a_new} is: 
\begin{equation}
a_m(t) = a_m(0) e^{\int_{t-\delta_m(t)}^{t} d\tau K_m(\bm{s}(\tau))}\;. \label{amsolv}
\end{equation}
Clearly, the case $\delta_m(t) = t$ corresponds to \eqref{dyn_a}, while $\delta_m(t) = 0$ recovers the case of constant $a_m$'s which corresponds to the naive energy minimization case.  
One approach to choosing  $\delta_m(t) $ is setting it to a small value  initially and doubling it every time the dynamics is stuck or hits an upper threshold ( set,  e.g., by a maximum allowed voltage value). This typically only requires a few iterations. Other  delay functions are being investigated.
It is important to note that  the decrease of satisfied clause's associated $a_m$ due to this time-delayed form relatively reduces the clause's weight in \eqref{dyn_s1}, thus increases other clauses' weights in the focused search space, enhancing the driving capability of unsatisfied clauses.
%Next we present several analog circuit designs that implement the above equations, with some variations.

%It is clear from Equ.~(\ref{newamequ}) that due to Equ.~(\ref{newaminitial}) we will always have $a_m(t)>0$ for any $t$. However, the $a_m$s do not have to be only increasing, as seen from Equ.~(\ref{dyn_a_new}). As a matter of fact, if for example at $t$, clause $K_m(\textbf{s}(t))$ is practically satisfied (very small), the corresponding $a_m$ will keep decreasing to a constant value (because $\delta_m(t)$ earlier $K_m$ was larger).

%\input{framework}
% !TEX root = main.tex
\section{System Design}
\label{sec:framework}
%Here  we introduce the detailed design of our analog SAT solver system, \mssat, built on the CTDS theory in Section~\ref{sec:background}. 
%\subsection{Overview of MS-SAT}

In this section, we present AC-SAT, our proposed analog SAT solver circuit based on the CTDS theory in Sec. \ref{sec:background}.  Though it is possible to
implement the CTDS equations digitally, the hardware would be much more costly in terms of area, power and performance. Thus we opt for an analog implementation, which also bears affinity with the operations in the CTDS. Our circuit design aims to keep the hardware solver configurable and modular while keeping the circuit simple and power efficient. These considerations require careful design of the overall architecture and some modifications to the algorithm itself, which will be elaborated later in the paper.

\begin{figure}[!t]
	\centering
	\includegraphics[width=0.5\textwidth]{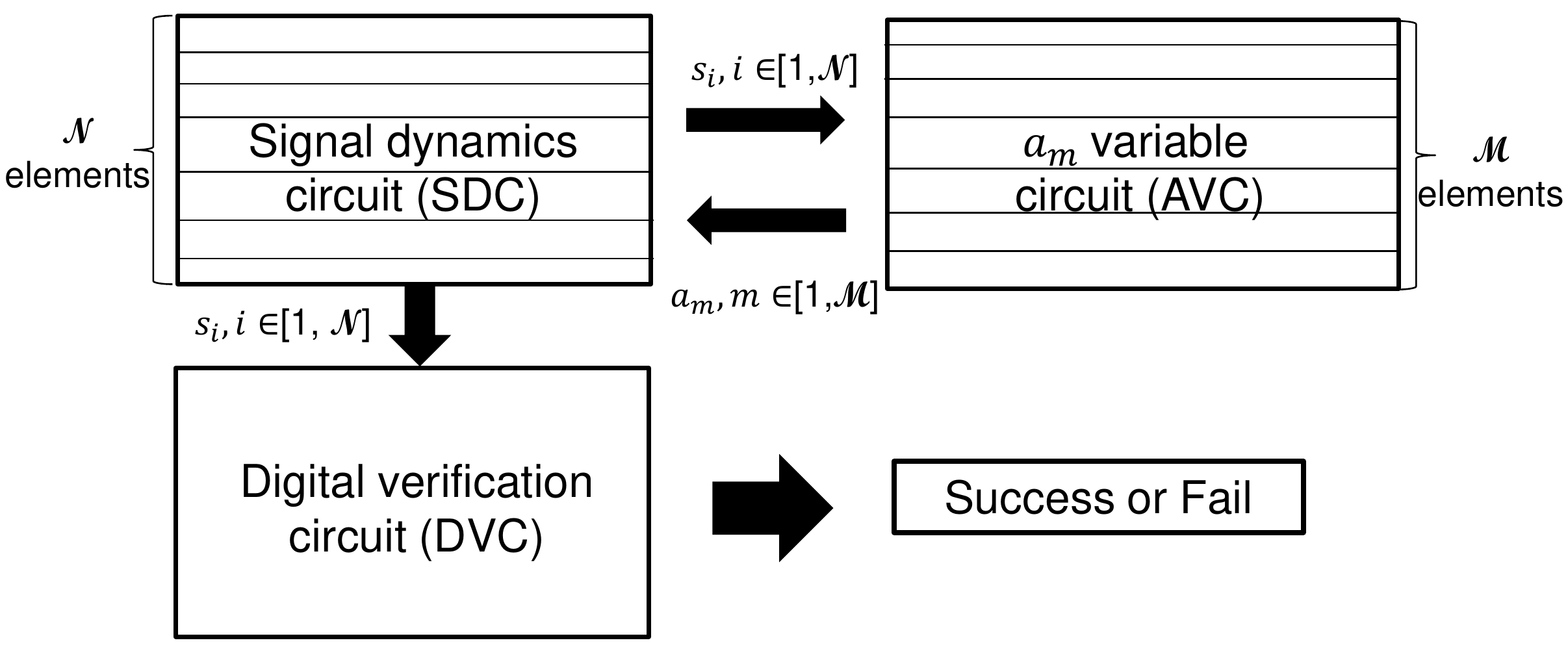}
	\caption{High-level block diagram of \mssat. The \sd contains  \cn elements while the \av contains  \cm elements. It can solve $k$-SAT problem instances with up to  \cn variables and  \cm clauses.}
	\label{fig:framework}
	%		\vspace*{-2ex}
\end{figure}

Fig. ~\ref{fig:framework}  shows the high-level  block diagram of \mssat.  It consists of three main components:  signal dynamics  circuit (\sd) which implements the dynamics of variable signals $s_i$'s in \eqref{dyn_s1}, auxiliary variable circuit (\av) which implements the dynamics of auxiliary variables $a_m$'s in \eqref{dyn_a}, and digital verification circuit (\dv) which checks whether all the clauses have been satisfied and outputs the satisfied assignments of variables.  
The \av  contains \cm identical elements, each of which receives the relevant
$s_i$'s signals from the \sd  as inputs, and generates $a_m$ $(m\in [1, M])$  (where $M \leq \cm$) variables as outputs. The \sd, containing  \cn identical elements,  in turn receives $a_m$'s as feedback from the \av  and evolves the $s_i$ $(i \in [1,N])$ signals (with $N \leq \cn$), accordingly. The \sd  outputs the analog values of $s_i$'s to the \av  and the digital version of $s_i$'s to the \dv. Based on the digital values of $s_i$'s, the \dv  
%uses a gate array programmed according to the problem clauses as well as an $AND$ tree to 
determines whether a solution to the given SAT problem has been found at that time. 

The given  block diagram can solve any $k$-SAT problem with $N$ or up to \cn Boolean variables and $M$ or up to \cm clauses.
However, a naive and direct implementation of the dynamical equations \eqref{dyn_a}, \eqref{dyn_s1} and \eqref{Dm1} would incur large hardware costs. Instead, here we present 
implementations of the  \sd and \av circuits that are much more resource-efficient than the direct approach.
Below, we elaborate the design of the three circuit components using the 3-SAT problem (i.e., three non-zero $c_{m,j}$'s for each clause) as an example. \mssat for any $k$-SAT problem can be designed following the same principle. 

%The two circuits contain arrays of analog elements that implement the dynamics specified by Eq.(\ref{dyn_a}) and Equ.(\ref{dyn_s}), and  the verification digital circuit receives the digital $s_i$ signals transfered from analog signals in signal dynamics circuit, feeds them into the gate arrays programmed based on the problem clauses and finally through $AND$ tree, gets the boolean flag indicating whether the $s_i$ signals have formed the solution.

\subsection{Signal Dynamics Circuit}
\label{sec:SD}

The \sd  contains an array of analog elements that realize the dynamics specified by \eqref{dyn_s1} and \eqref{Dm1}. Though it is possible to implement the multiplications  and voltage controlled current source (VCCS) in \eqref{dyn_s1} and \eqref{Dm1}  straightforwardly based on operational amplifiers, such implementations can be rather costly. We introduce several novel circuit design ideas to implement the dynamics in (5) and (6). We will show that the accuracy of the circuit is sufficient for the type of dynamical systems being considered here.

\begin{figure}[!t]
	\centering
	\includegraphics[width=0.4\textwidth]{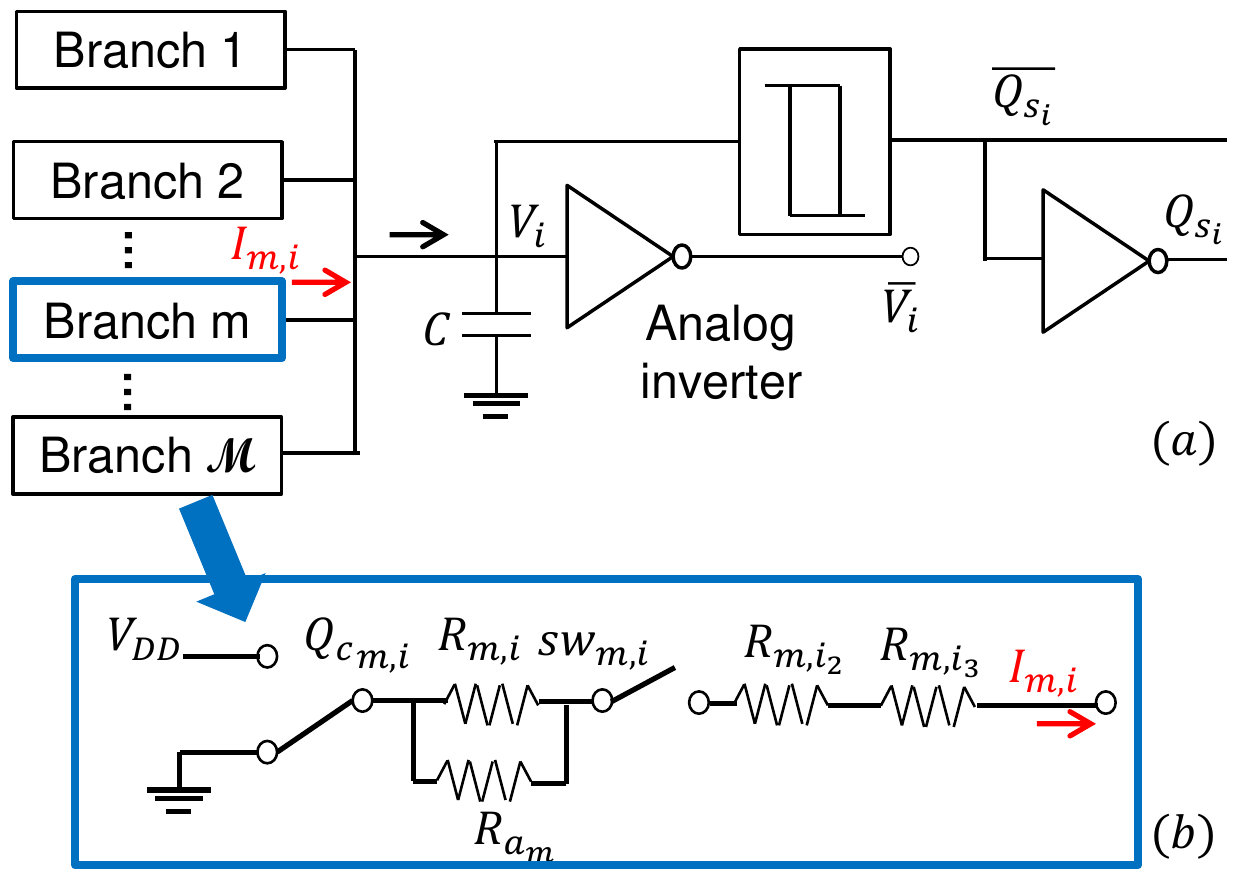}
	\caption{(a) Design of one array element in the \sd. (b) Detailed conceptual design of the Branch block.} %$N$ array elements are needed for $N$ Boolean variables in a 3-SAT problem.}
	\label{fig:xbranch}
	%		\vspace*{-2ex}
\end{figure}

Given a 3-SAT problem with $N$ variables, the \sd  enables an array of $N$ analog elements, referred to as $s_i$ element, for evaluating the $s_i$ ($i=1,\ldots,N$) signals. Fig.~\ref{fig:xbranch}(a) shows the conceptual design of the $s_i$ element that realizes \eqref{dyn_s1}. The $s_i$ element contains a capacitor, $C$, connected to the $M$ Branch blocks (where $M$ is the total number of clauses in the 3-SAT problem), an analog inverter, an inverted Schmitt trigger and a digital inverter. The voltage across capacitor $C$, i.e., $V_i$, and the output of the analog inverter, $\overline{V_i}$, represent the analog value of signal $s_i$ and $-s_i$, respectively.   Signal $s_i \in [-1,1]$ (resp., $-s_i \in [1,-1]$) is mapped to $V_i \in  [G\!N\!D, V_{DD}]$ (resp., $\overline{V_i} \in [V_{DD}, G\!N\!D]$). The inverted Schmitt trigger and the digital inverter output the  digital versions of $-s_i$ and $s_i$, denoted by $\overline{Q_{s_i}}$ and $Q_{s_i}$, (i.e., taking on values of either $G\!N\!D$ or $V_{DD}$).

To see why the $s_i$ element in Fig.~\ref{fig:xbranch}(a) can be used to evaluate \eqref{dyn_s1}, let us
denote the current from each of the Branch block as $I_{m,i}$. Then we have
\begin{equation}
\label{vi}
C\frac{dV_{i}}{dt} = \sum_{m=1}^{M} I_{m,i} 
\end{equation}
Comparing \eqref{dyn_s1} with \eqref{vi}, we see that the $s_i$ element in Fig.~\ref{fig:xbranch}(a) precisely realizes \eqref{dyn_s1} if we have
\begin{equation}
\label{im}
I_{m,i} =  C a_{m} D_{m,i} 
\end{equation}
In order to design a Branch block to satisfy \eqref{im}, we first  make some observations  related to the $D_{m,i}$  quantities. In a 3-SAT problem, there are only three non-zero $c_{m,j}$'s in each $C_m$ clause. Let us denote them as $c_{m,i}$, $c_{m,i_2}$, and $c_{m,i_3}$. Then $D_{m,i}$ in \eqref{dyn_s1} can be shown to have the following form:
\begin{multline}
\label{Dm2}
D_{m,i} =  2^{-2} \times \\ 
c_{m,i} (1-c_{m,i}s_{i}) (1-c_{m,i_2}s_{i_2})^2 (1-c_{m,i_3}s_{i_3})^2 = \\
\begin{cases}
2^{-2}(+1-s_{i}) (1-c_{m,i_2}s_{i_2})^2 (1-c_{m,i_3}s_{i_3})^2 & \!\!\! \mbox{if }c_{m,i}=1 \\
0& \!\!\! \mbox{if }c_{m,i}=0 \\
2^{-2}(-1-s_{i}) (1-c_{m,i_2}s_{i_2})^2 (1-c_{m,i_3}s_{i_3})^2 & \!\!\! \mbox{if }c_{m,i}=-1
\end{cases} 
\end{multline}
Referring to (\ref{Dm2}), one can readily see that $D_{m,i}$ has the following properties:
\begin{itemize}
	\item If any of $s_{i}$, $s_{i_2}$ and $s_{i_3}$ is satisfied, i.e., reaches $1$ or $-1$ (indicating $x_i$ is either TRUE or FALSE), $D_{m,i}$ becomes zero. %For example, if $c_{m,i}$ is one (i.e. $i$th variable $x_i$ appears in $m$th clause), then $s_{i}=1$ results in $D_{m,i}=0$. 
	According to (\ref{dyn_s1}), a zero $D_{m,i}$ means that clause $C_m$ has no impact on the variation of $s_{i}$. On the other hand, when none of the three variables is satisfied, but one of them gets closer to being satisfied, the magnitude of $D_{m,i}$ reduces,  again.
	%\vspace{-1.ex}
	\item The sign of $D_{m,i}$ is the same as that of $c_{m,i}$ since $(1-s_{i})$ cannot be negative. If $c_{m,i} = 1$ (resp., $-1$), the contribution of $D_{m,i}$ to $\frac{ds_{i}}{dt}$ is positive (resp., negative), i.e., it tries to push $s_{i}$ toward $+1$ (resp., $-1$).
\end{itemize}

Based on the above observations, let us examine the conceptual design of the Branch block in Fig.~\ref{fig:xbranch}(b). Specifically, the Branch block contains 
two switches and  four tunable resistive elements. (The resistive elements here are used to simplify the drawing,  and the details about their design will be described  later.)  Here $R_{m,i}$ represents the resistive element associated with $s_i$, while $R_{m,i_2}$ and $R_{m,i_3}$ represent the resistive elements associated with the other two signals $s_{i_2}$ and $s_{i_3}$ in \eqref{Dm2}. {\hu $R_{a_m}$ represents the resistive element associated with auxiliary variable $a_m$.}
One switch is controlled by signal $sw_{m,i}$, which is left open if $c_{m,i} =0$ (indicating that $x_i$ does not appear in  clause $C_m$), and is closed otherwise.
The other switch is controlled by $Q_{c_{m,i}}$, the digital version of $c_{m,i}$. If $c_{m,i} =1$ (resp., $-1$), indicating that  $x_i$ (resp., $\overline{x_i}$) is present in the clause, the switch controlled by $Q_{c_{m,i}}$ connects to $V_{DD}$ (resp., $G\!N\!D$).  It can be readily seen that
\begin{equation}
\label{eq:branch}
I_{m,i} = 
\begin{cases}
(V_{DD} - V_{i}) / (R_{m,i}||R_{a_m} + R_{m,i_2} + R_{m,i_3})   & \mbox{if }c_{m,i}=1\\
0  & \mbox{if }c_{m,i}=0 \\
(G\!N\!D-V_{i}) / (R_{m,i} || R_{a_m} + R_{m,i_2} + R_{m,i_3})  & \mbox{if }c_{m,i}=-1 \\
\end{cases} 
\end{equation}    
If the values of $R_{a_m}, R_{m,i}, R_{m,i_2}$ and $R_{m,i_3}$ are chosen properly, the $I_{m,i}$ value derived from the Branch block would have the same properties as identified for $D_{m,i}$ above.

% \begin{figure}[!t]
% 	\centering
% 	\includegraphics[width=0.5\textwidth]{branch_amr1rii2_1}
% 	\caption{Actual circuit implementation of the conceptual Branch block design in Fig.~\ref{fig:xbranch}(b). The red box (lower part) includes the implementation of the switch as well as $R_a$ and $R_i$. The green box (on the right) includes the implementation for $R_{i_2}$ and $R_{i_3}$.}
% 	\label{fig:branch}
%		\vspace*{-2ex}
% \end{figure}

\begin{figure}[!t]
	\centering
	\includegraphics[width=0.5\textwidth]{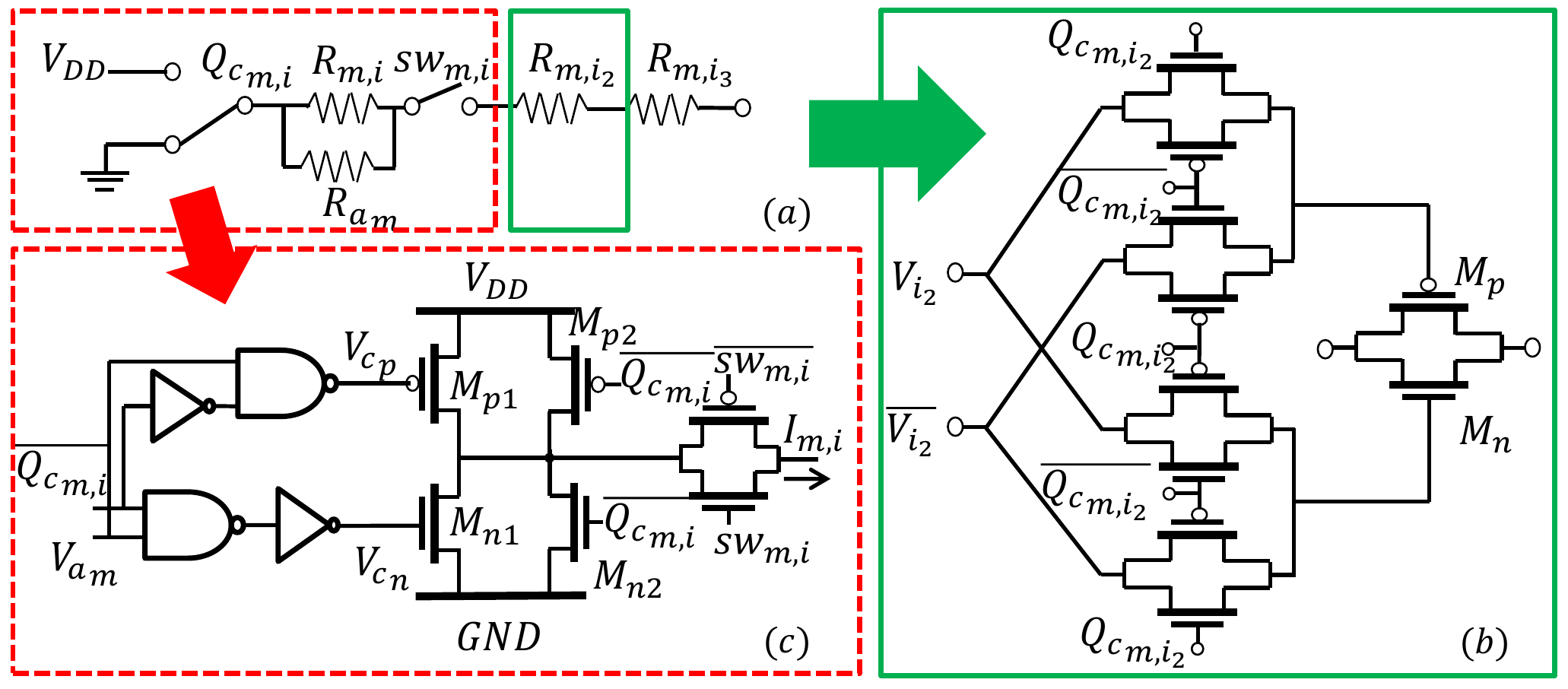}
	\caption{Detailed design of the Branch block in Fig.~\ref{fig:xbranch}(b).  (a) The conceptual design of the Branch block. %The red box  includes the implementation of the switch as well as $R_a$ and $R_i$. 
		(b) Circuit implementation for $R_{m,i_2}$ and $R_{m,i_3}$ (elements in the green box in Fig.~\ref{fig:branch}(a)). 
		(c) Circuit implementation for the switch as well as $R_{a_m}$ and $R_{m,i}$ (elements in the red box in Fig.~\ref{fig:branch}(a)).}
	\label{fig:branch}
	%  	\vspace*{-2ex}
\end{figure}

\begin{figure}[!t]
	\centering
	\includegraphics[width=0.5\textwidth]{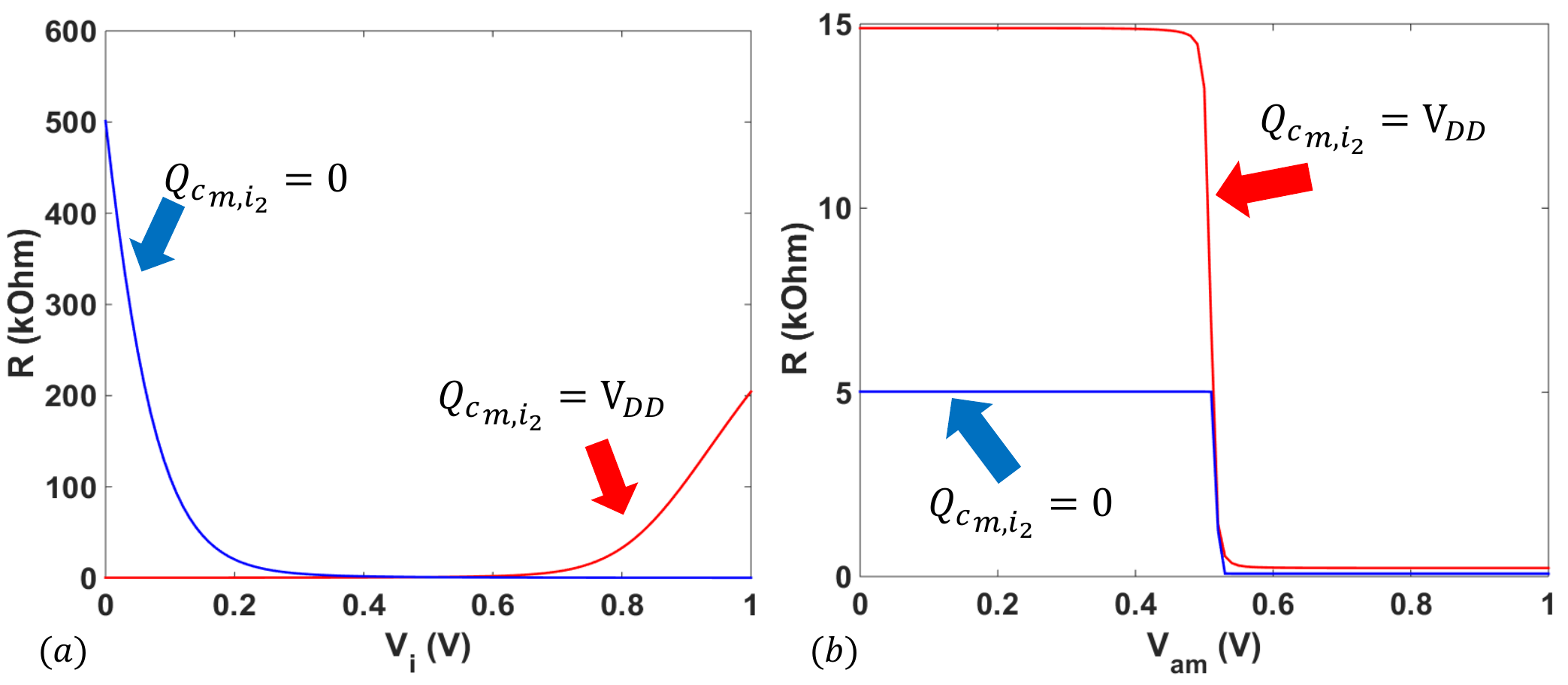}
	\caption{SPICE simulation results depicting (a) the  value of resistor $R_{m,i_2}$ or $R_{m,i_3}$ in Fig.~\ref{fig:branch}(a)  as function of $V_i$. (b) the value of resistor $R_{m,i} || R_{a_m}$ in Fig.~\ref{fig:branch}(a)  as function of $V_{a_m}$.}
	\label{fig:waves}
	%   	\vspace*{-2ex}
\end{figure}

The actual realization of the four resistive elements in Fig.~\ref{fig:xbranch}(b) is given in Fig.~\ref{fig:branch}.
The implementation of $R_{m,i_2}$  and that of $R_{m,i_3}$ are the same and the one for $R_{m,i_2}$ is shown  in Fig.~\ref{fig:branch}(b). Consider the $R_{m,i_2}$ block. The two terminals of the transmission gate formed  by transistor $M_p$ and $M_n$ correspond to the terminals of $R_{m,i_2}$. The gate terminals of  $M_p$ and $M_n$ are connected to $V_{i_2}$ and $\overline{V_{i_2}}$ via four additional transmission gates  controlled by $Q_{c_{m,i_2}}$ and $\overline{Q_{c_{m,i_2}}}$. It can be derived that this realization of $R_{m,i_2}$ exhibits the desired properties outlined above for $D_{m,i}$  in \eqref{Dm2}. The SPICE simulation results depicting the relationship between the resistance value and $V_i$ are given in Fig.~\ref{fig:waves}(a). For example, assuming that $c_{m,i_2}$ is 1, i.e. $Q_{c_{m,i_2}}=V_{DD}$  (corresponding to the red line in Fig.~\ref{fig:waves}(a)), the gates of $M_n$ and $M_p$ are connected to $\overline{V_{i_2}}$ and $V_{i_2}$, respectively. If variable $s_{i_2}$ is satisfied, i.e. $V_{i_2}$ is close to $V_{DD}$ (where $V_{DD}=1$ $V$ in Fig.~\ref{fig:waves}(a)) and $\overline{V_{i_2}}$ is close to $G\!N\!D$, then both $M_n$ and $M_p$ are OFF, $R_{m,i_2}$ has a very large value  (around 200 k$\Omega$) and $I_{m,i}$ is close to zero. This means that clause $C_m$ has no impact on the variation of $s_{i}$ which is exactly the desired behavior. On the other hand, if $s_{i_2}$ is not satisfied,  as it gets closer to its target  (i.e., $+1$), the magnitude of $I_{m,i}$ reduces because $R_{m,i_2}$ increases  as can be seen by the increase in the resistance value as $V_i$ gets close to $V_{DD}$.
The blue line in Fig.~\ref{fig:waves}(a) corresponds to the case where $c_{m,i_2} = -1$ and its behavior can be explained in the same way as above.

%{\cg assume that $c_{m,i_2}$ is 1 (resp., $-1$), i.e. $Q_{c_{m,i_2}}=V_{DD}$ (resp., $0$), corresponding to the red line (resp., the blue line) in Fig.~\ref{fig:waves}(a))}. The gates of $M_n$ and $M_p$ are connected to $\overline{V_{i_2}}$ and $V_{i_2}$, respectively. If variable $s_{i_2}$ is satisfied, i.e. $V_{i_2}$ is close to $V_{DD}$ {\cg (where $V_{DD}=1 V$ in Fig.~\ref{fig:waves}(a)} and $\overline{V_{i_2}}$ is close to $G\!N\!D$, then both $M_n$ and $M_p$ are OFF, $R_{i_2}$ has a very large value {\cg (around $200 k\Ohm)$}and $I_{m,i}$ is close to zero. This means that clause $C_m$ has no impact on the variation of $s_{i}$ which is exactly the desired behavior. On the other hand, if $s_{i_2}$ is not satisfied,  as it gets closer to its target (either $+1$ or $-1$), the magnitude of $I_{m,i}$ reduces because $R_{i_2}$ increases {\cg as can be seen by the increase in the resistance value (either as $V_i$ gets close to $V_{DD}$ or $0$). Note that 

%Should you use V_{a_m}} instead of just a_m? Should \overline{c_mi} be $V_{DD} \codt  \overline{c_mi}
The circuit block for implementing the switch controlled by $Q_{c_{m,i}}$ and the two resistive elements $R_{m,i}$ and $R_{a_m}$ is shown in  Fig.~\ref{fig:branch}(c). 
%The circuit also incorporates the $a_m$ term in \eqref{dyn_s1}.
The gates of transistor $M_{p1}$ and $M_{p2}$ (resp., $M_{n1}$ and $M_{n2}$) control the connection to $V_{DD}$ (resp., $G\!N\!D$). One of the gates connects directly to $\overline{Q_{c_{m,i}}}$ (representing the negated $c_{m,i}$ signal) while the other is controlled by 
\begin{equation}
\label{cp}
V_{c_{p}} =  \overline{V_{a_{m}}Q_{c_{m,i}}}
\hspace{2em} \mbox{resp.,} \hspace*{1em}
V_{c_{n}} =  V_{a_{m}} \overline{Q_{c_{m,i}}}
\end{equation}
%{\cg (SH) 
Note that though $V_{a_m}$ in \eqref{cp} (as input to the two  \texttt{NAND} gates) seems to be treated as a digital signal, it actually remains as an analog signal while the \texttt{NAND} gates and inverters operate  in the linear $V_{in}-V_{out}$ region to produce analog outputs as desired. 
%(Thus it is possible that these gates consume static power in the process, however, at a small amount level since the transistor sizes are selected small.)
The SPICE simulation results in Fig.~\ref{fig:waves}(b) indicate that the block realizes the switch function due to $c_{m,i}$ as well as $(+1-s_{i})$ and $(-1-s_{i})$ in \eqref{Dm2}, and incorporates the $a_m$ term in \eqref{dyn_s1}.  Consider the case that $c_{m,i}=-1$, i.e. $Q_{c_{m,i}}=0$. Then $V_{c_p}=V_{DD}$ and $V_{c_n}=V_{a_m}$, which means that the  block  in Fig.~\ref{fig:branch}(c) is connected to $G\!N\!D$ and the current flowing through the block is dependent on the voltage representing $a_m$, $V_{a_m}$.   The blue line in Fig.~\ref{fig:waves}(b) shows the equivalent resistance value of the block versus $V_{a_m}$. As $V_{a_m}$ gets larger, $M_{n1}$ exhibits smaller resistance, resulting in larger impact of $a_m$ on the current flowing through the  block. Note that initially $V_{a_m}$ is very small, thus the right two transistors $M_{p2}$ and $M_{n2}$ which are of smaller sizes than $M_{p1}$ and $M_{n1}$ are employed to ensure proper current flow  as well as serve as a current boost.  The red line in Fig.~\ref{fig:waves}(b) corresponds to the case where $c_{m,i}=1$, i.e. $Q_{c_{m,i}}=V_{DD}$.

\begin{figure}[!t]
	\centering
	\includegraphics[width=0.5\textwidth]{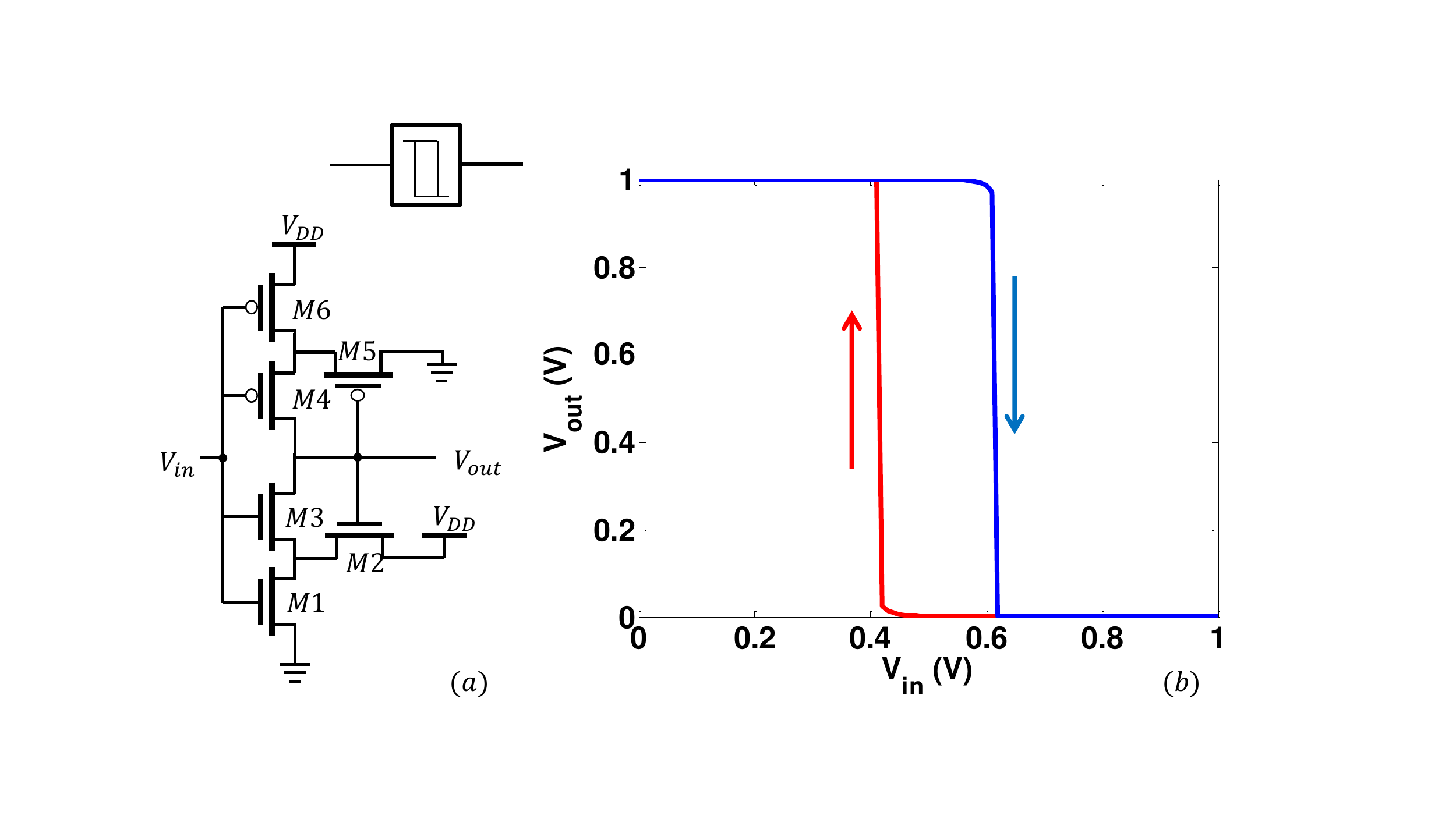}
	\caption{Schmitt-trigger inverter and its transfer characteristic.}
	\label{fig:schmitttrigger}
	%		\textbf{\vspace}*{-2ex}
\end{figure}

The \sd also converts the analog signals, $V_i$'s, to digital signals, $Q_{s_i}$'s, via an inverted Schmitt trigger. The inverted Schmitt trigger circuit is shown in Fig.~\ref{fig:schmitttrigger}(a). The digital signals are then sent to the \dv  to check if a solution has been found. The inverted Schmitt trigger circuit exhibits hysteresis in its transfer curve as seen from the simulation result in Fig.~\ref{fig:schmitttrigger}(b), hence can perform analog-digital conversion with minimal noise impact. Putting all the above discussions together, one can conclude that the \sd  correctly implements the system dynamics defined by \eqref{dyn_s}.

\subsection{Auxiliary Variable Circuits}
\label{sec:AVDV}
%At the moment a good method to incorporate $a_{m}$ in the circuit is not developed. Hence a extra resistance in each switch which is controlled with $a_{m}$ is used in circuit simulation. Introduction of R makes the circuit slow and We have found that there is a possibility that the circuit gets trapped. It seems that better ideas both at the circuit level and algorithm level are needed to eliminate the need for $a_{m}$ and replace it with some form of random resetting of the Vi's or something similar.
As pointed out in Section~\ref{sec:background}, the auxiliary variables, $a_m$'s as defined in \eqref{dyn_a}, are used to help avoid the gradient descent search being stuck in non-solution attractors. The $a_m$ signal follows an exponential growth driven by the level of non-satisfiability in clause $C_m$. A direct way to implement an exponential function is through an operational amplifier (op-amp), which we present below.
Note that we have realized the analog version of equation \eqref{dyn_a} in a resource-efficient manner, similar to the implementation in \dv, to avoid costly multiplications and VCCS implementations.

\begin{figure}[!t]
	\centering
	\includegraphics[width=0.4\textwidth]{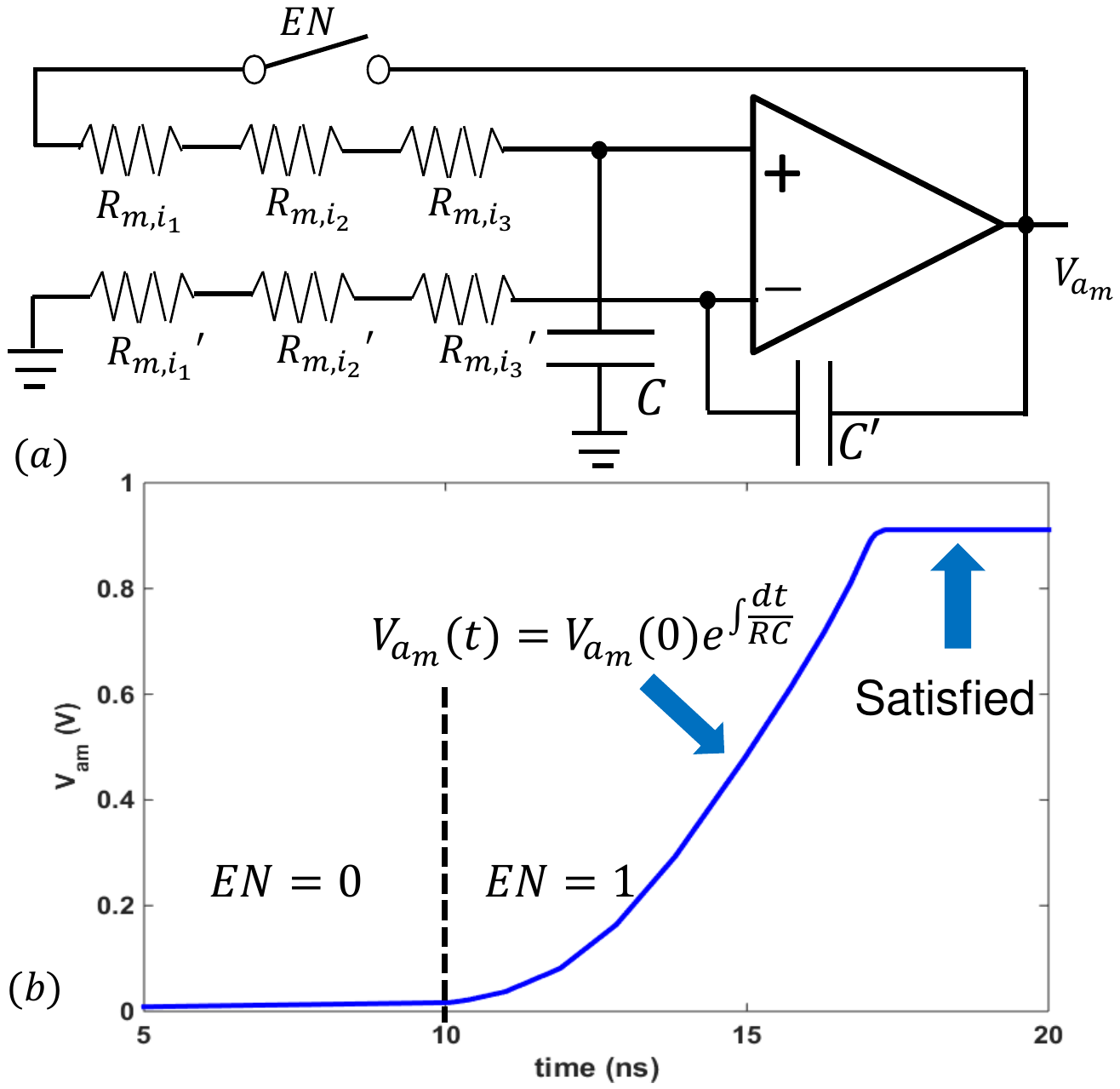}
	\caption{ (a) Conceptual design of one element in the \av. Actual realization of the resistive element is similar to the circuit in Fig.~\ref{fig:branch}(b). (b) SPICE simulation result depicting the waveform of $V_{a_m}$ vs. time.  $V_{a_m}(0)$ represents the initial value of $V_{a_m}$.} %A total of $M$ such elements are needed to generate all the $a_m$'s.}
	\label{fig:expona}
	%		\vspace*{-2ex}
\end{figure}

The \av  contains an array of \cm $a_m$ elements where \cm is the maximum number of clauses in a given problem that the \av can handle.  Fig.~\ref{fig:expona} illustrates the conceptual design of the $a_m$ element, similar to a non-inverting integrator. Here, the value of $a_m$ (for clause $C_m$) is represented by the voltage at the output of the op-amp,  i.e., $V_{a_m}$.  
Resistive elements $R_{m,i_1}$, $R_{m,i_2}$ and $R_{m,i_3}$ {\hu are associated with the three signals in $a_m$'s clause while $R_{m,i_1}'$, $R_{m,i_2}'$ and $R_{m,i_3}'$ are identical to  $R_{m,i_1}$, $R_{m,i_2}$ and $R_{m,i_3}$, respectively.} The two capacitors, $C$ and $C'$, have identical values as well.  Together with the resistive elements, they control the speed of $V_{a_m}$ growth. The switch controlled by $E\!N$ is realized by a transmission gate to control the start of  the $a_m$ element.
The first order differential equation of $V_{a_m}$ can be written as:
\begin{equation}
\label{eq:expona}
C\frac{dV_{a_m}}{dt} =V_{a_m}/(R_{m,i_1}+R_{m,i_2}+R_{m,i_3}).
\end{equation}
%$R_{i_1}$ (resp., $R_{i_1}'$), $R_{i_2}$ (resp., $R_{i_2}'$), $R_{i_3}$ (resp., $R_{i_3}'$)
The R's in Fig.~\ref{fig:expona}(a) are tunable resistive elements implemented by transmission gates which have similar circuit topology to that shown in the green box of Fig.~\ref{fig:branch}. For example, for $R_{m,i_1}$ and $R_{m,i_1}'$, the transmission gates ($M_p$ and $M_n$) are controlled, via four other transmission gates, by analog signals $V_{i_1}$ and $V_{i_1}'$, representing $x_i$'s presence in clause $C_m$. The other two pairs of $R$'s are designed in the same way. If any of the three variables in clause $C_m$ is satisfied, the corresponding $V_i$ turns off the respective transmission gate and cut off the current paths from op-amp's inverting input  to  ground and from non-inverting input to $V_{a_m}$. 

The circuit in Fig.~\ref{fig:expona}(a) exactly realizes the exponential growth specified in \eqref{dyn_a}  up to an upper bound  on $V_{a_m}$, i.e. the op-amp's supply voltage.  Fig.~\ref{fig:expona}(b) plots an example $V_{a_m}$ value growth with time before and after associated signals $s_i$'s get satisfied. After $E\!N$ is set to $1$ (i.e., the switch is closed), $V_{a_m}$ starts to grow exponentially, following the differential equation in \eqref{eq:expona}. According to Fig.~\ref{fig:waves}(b), as $V_{a_m}$ increases, the resistant value of $R_{m,i} || R_{a_m}$ drops down, leading to a larger current $I_{m,i}$ in the corresponding Branch block in Fig.~\ref{fig:xbranch}(b), which is consistent with \eqref{im}. This current, together with other currents that are associated with $V_i$ in Fig.~\ref{fig:xbranch}(a), contributes to the variation of $V_i$ which is specified in \eqref{vi}. There are two cases that may stop the evolution of $V_{a_m}$:
(i) As stated above, if any one of the three analog signals in clause $C_m$ is satisfied,  the current paths in Fig.~\ref{fig:expona}(a) is cut off, and $V_{a_m}$ stops at a certain voltage. This indicates that $V_{a_m}$ has finished its utility as an auxiliary variable to drive the corresponding clause to the satisfied state; (ii)
If $V_{a_m}$ reaches its upper bound before any of the three variables in the corresponding clause is satisfied, the circuit  stops evolving since the  $V_{a_m}$ value is unable to drive this yet unsatisfied clause any more. This impacts the effectiveness of avoiding being stuck in a non-solution attractor during the gradient descent search process. % (circuit diffusion process). 
The upper bound on $V_{a_m}$  imposes a physical limitation  on the hardware realization of the CTDS theory\footnote{As discussed in Sec.~\ref{sec:background}, Eqs. \eqref{dyn_s}-\eqref{dyn_a} are not unique, and the effect of the maximum voltage limitation depends on the equations themselves. For example, Sec.~\ref{sec:background} introduces an alternative, delay-based formulation for $a_m$ in \eqref{dyn_a_new} which allows $a_m$ to decrease when the corresponding clause is satisfied. This delay-based formulation of $a_m$  postpones reaching the $a_m$ upper bound. The implementation of \eqref{dyn_a_new}  for the op-amp based approach is currently under development.}.

Although the \av  design given in Fig.~\ref{fig:expona} realizes the exponential growth, it requires \cm op-amps, resulting in a large amount of area and power consumption.   There exist other ways to achieve exponential signal growth, e.g., circuits with positive feedback often have exponential growth in certain ranges. 
%{\cg Zoltan: add a bit more about the potential effects of this function vs. the exponential function?}
Besides the exponential growth implementation, we will introduce alternative \av  designs in the next subsection.

\subsection{Digital Verification and Interface Circuits}
\label{sec:dvio}
\begin{figure}[!t]
	\centering
	\includegraphics[width=0.5\textwidth]{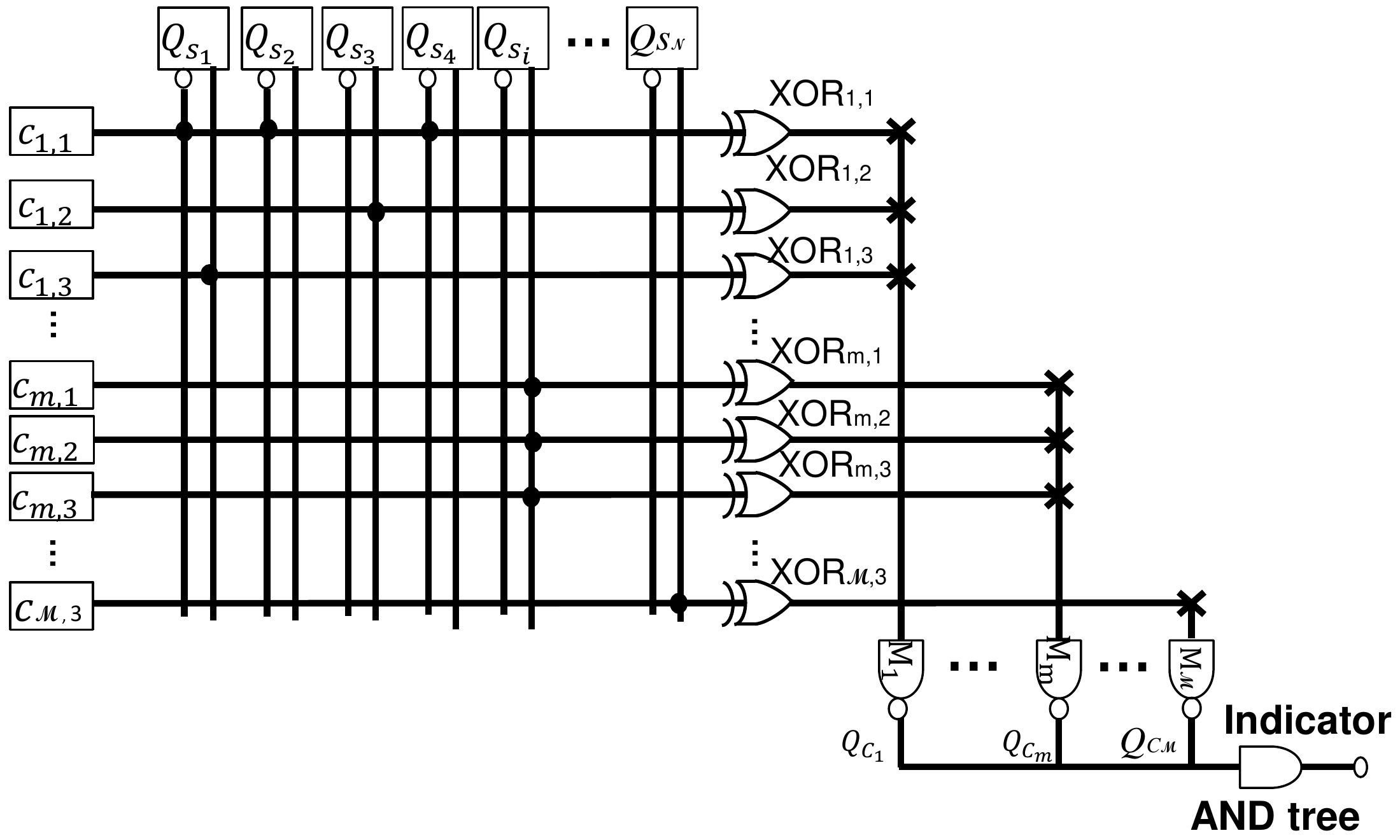}
	\caption{Schematic of the \dv.}
	\label{fig:dv}
	%		\vspace*{-2ex}
\end{figure}

The goal of the \dv  is to determine if a solution (the set of $s_i$'s) to the given problem has been found  within a user specified time bound. 
%One important observation of the CTDS for solving a $k$-SAT problem is that once the sign of $s_i$ is switched from one to the opposite one, $s_i$ can be considered as satisfied. {\bf Xunzhao: check if your circuit exploits this property.} Following this principle,
The \dv  is implemented readily through the use of an array  of   $3M$ \texttt{XOR} gates and an array of $M$ \texttt{NAND} gates as shown in Fig.~\ref{fig:dv}. The input to the \dv  is the digital representation of $s_i$'s and $-s_i$'s, i.e., $Q_{s_i}$ and $\overline{Q_{s_i}}$, from the \sd. Each \texttt{NAND} gate corresponds to a clause and its inputs correspond to the literals present in the clause. Note that in the \dv, we only include those $c_{m,i}$'s whose values are $+1$ (represented by logic signal ``1'') and $-1$ (represented by logic ``0'').  The outputs of the \dv are analog values $Q_{C_m}$, for clauses $C_m$, and Indicator, which is set to 1 if the circuit finds a solution, otherwise it remains at 0. The \dv is an asynchronous circuit, and the output of the \dv constantly records whether a solution is found or not. By setting a  time bound $T$, the \dv regards the problems whose solutions are found within $T$ as satisfiable problems, the rest are considered either unsatisfiable or unsatisfiable within the alloted time. Note that for problem instances where no solutions are found in the given time bound, our approach does not provide a formal proof of unsatisfiability (as our algorithm is an incomplete algorithm). However, our solver is a MaxSAT solver, because it does not use any assumptions about the solvability of the formula and minimizes the number of unsatisfied clauses within the allotted resources or time. Theoretical analysis of the performance of the solver as a MaxSAT solver is out of the scope of this paper and will be presented elsewhere.

It is easy to see that all three components, \sd, \av, and \dv, are modular and programmable.  By modular, we mean that the basic elements in each circuit can be repeated for different problem sizes (i.e., number of variables, $N$, and number of clauses, $M$). By programmable, we mean that any $k$-SAT problem instance can be solved by the same \sd, \av and \dv implementation as long as the problem size is less than or equal to the hardware specification. 
%uses a gate array programmed according to the problem clauses as well as an $AND$ tree to 

%The two main blocks of the solver core process the signals $\textbf{s}$ and $a_m$s respectively and affect each other 
Below, we briefly describe the I/O interface between the CPU and \mssat.
\mssat is used as a co-processor, similarly to other reconfigurable co-processors (such as dynamically reconfigurable FPGAs). To facilitate configuration, AC-SAT can be augmented with an on-chip reconfiguration memory as well as a simple controller. Based on the problem description (given in the CNF), CPU writes to memory the configuration information. The controller then uses the memory contents to set the respective switches in the \sd, \av and \dv components. 

\begin{figure}[!t]
	\centering
	\includegraphics[width=0.45\textwidth]{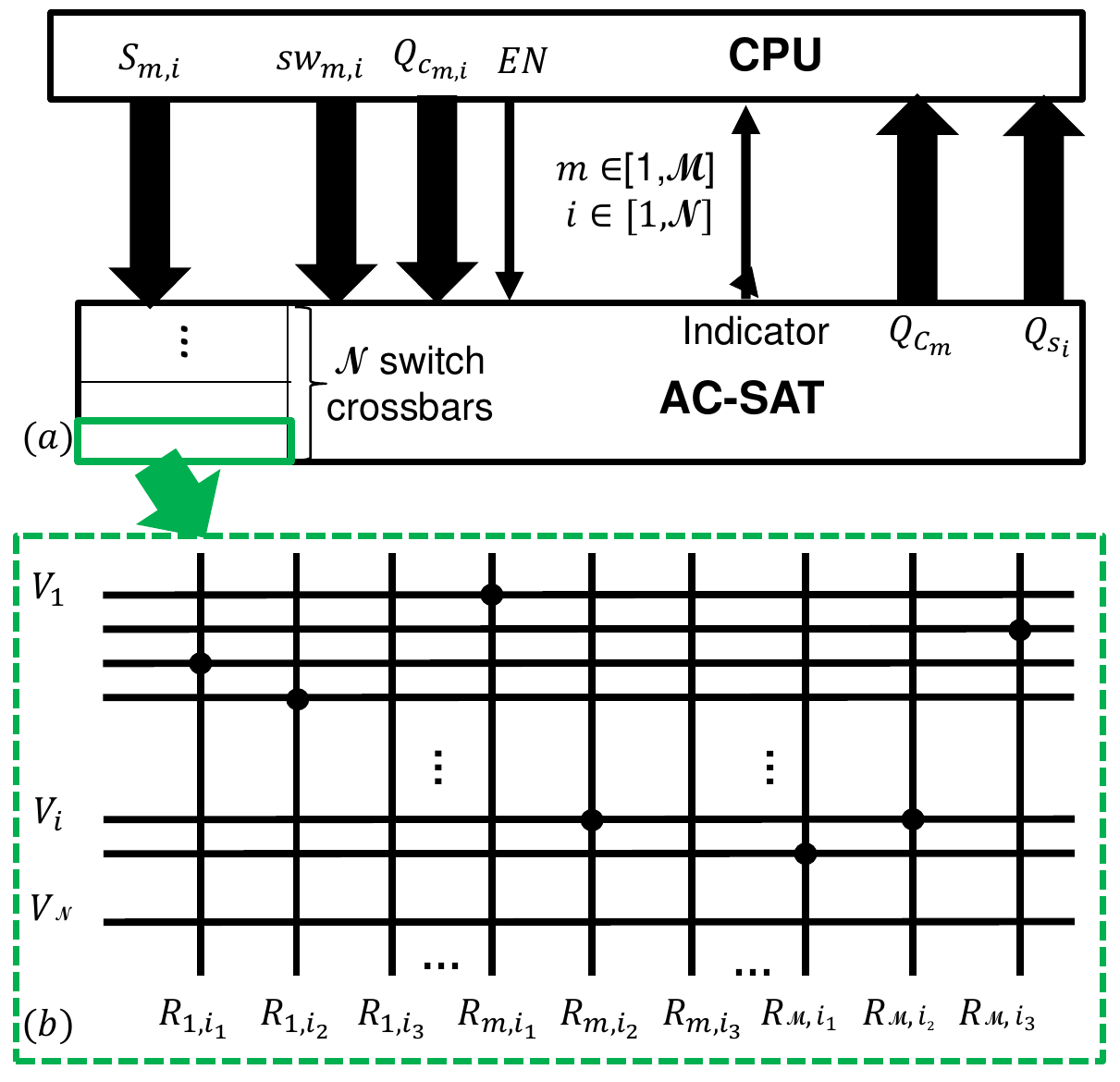}
	\caption{(a) Inputs and outputs of \mssat. (b) The structure of a switch crossbar. }
	\label{fig:io}
	%		\vspace*{-2ex}
\end{figure}

The main configuration information sent from CPU to AC-SAT includes the following: $E\!N$, $S_{m,i}$, $sw_{m,i}$ and $Q_{c_{m,i}}$. $E\!N$ activates \mssat.
%\sout{Based on the CNF of the problem instance, CPU set the values of $S_{m,k}$ $sw_{m,i} $ and $Q_{c_{m,i}}$.}
$S_{m,i}$ describes the appearance of variable signals in the corresponding clauses, e.g., $S_{m,i}=$1 (or 0) means that variable $s_i$ is (or is not) in the $m^{th}$ clause. $sw_{m,i}$ is used to deactivate unused Branch blocks in the \sd, e.g., $sw_{m,i}=0$ would deactivate the $m^{th}$ Branch block for variable $s_i$. 
%\sout{These signals, along with enable signal $EN$ are sent from CPU to \mssat.}
$Q_{c_{m,i}}$ is the digital version of $c_{m,i}$ (see Sec. \ref{sec:SD} for its definition).
\mssat receives the inputs from CPU, delivers $sw_{m,i}$ and $Q_{c_{m,i}}$ to the memory associated with the \sd component.
Since $S_{m,i}$ cannot be used directly by the resistive elements in the \dv and \av which require the internal variable signals $V_i$'s (see Fig. \ref{fig:branch} and Fig. \ref{fig:expona}(a)), we use \cn switch crossbars (Fig. \ref{fig:io}(b)) to accomplish the mapping from the variable signals $V_i$'s to each resistive element in the \sd and \av based on $S_{m,i}$ (i.e. $S_{m,i}$ is used to set the state of the corresponding switch). 
For output, \mssat indicates $Q_{s_i}$, $Q_{C_m}$ and Indicator from the \dv to CPU. $Q_{s_i}$ indicates the values of variable signals, and $Q_{C_m}$ indicates the states of all clauses (satisfied or not). All the inputs and outputs are digital signals. Once \mssat finds a solution, signal Indicator acts as an interrupt to CPU and CPU reads the \mssat outputs, i.e., $Q_{C_m}$ and $Q_{s_i}$. On the other hand, if the circuit runs out of time and no solution is found, \mssat outputs the results with minimum unsatisfiable clauses, but with Indicator = 0.

%\input{improved_designs}
% !TEX root = main.tex
\subsection{Alternative AVC  Designs}
\label{sec:improvedesign}

The op-amp based \av  described in Section~\ref{sec:AVDV} realizes an exponentially growing $a_m$ variable aiming to address hard %($\alpha > 4.2$, for 3-SAT) 
SAT problems (some SAT instances with constraint density $\alpha$=$M/N  \gtrsim  4.25$) within its physical limitation.
However, for application type SAT problems, i.e. which are not specially designed to be very hard, exponential growth for $a_m$ is not always necessary.
%However, this can be rather costly in terms of area and power. 
Below we describe two alternative circuit designs to implement an $a_m$ function that has a  $(1-\epsilon_2 e^{- q t})$-type growth to a saturation value. In the remainder we will refer to this version of $a_m$ growth as the ``simpler version". 

\begin{figure}[!t]
	\centering
	\includegraphics[width=0.35\textwidth]{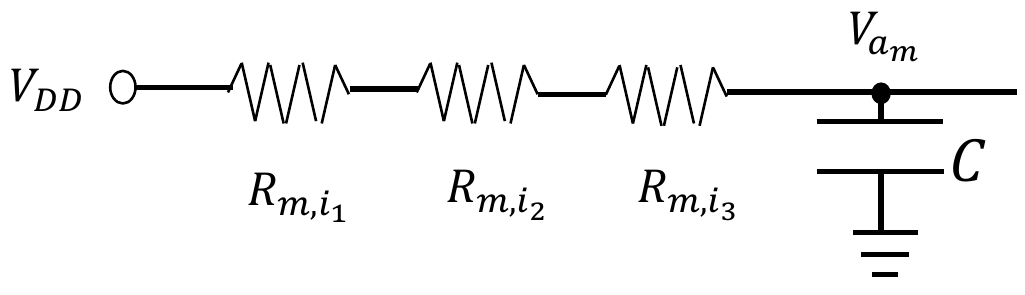}
	\caption{Conceptual design of the  \av element realizing the $(1-\epsilon_2 e^{- q t})$-type growth. Implementation of the resistive elements is similar to those in Fig.~\ref{fig:branch}. }%A total of $M$ such elements are needed to generate all the $a_m$'s.}
	\label{fig:am-simple}
	%		\vspace*{-2ex}
\end{figure}

Fig.~\ref{fig:am-simple} depicts the conceptual design of the $a_m$ element realizing the simpler $a_m$ growth, where 
capacitor $C$ is charged to $V_{DD}$ through three tunable resistors. The first order differential equation that governs $V_{a_m}$ can be written as 
\begin{equation}
\label{vam}
C\frac{dV_{a_m}}{dt} = (V_{DD}-V_{a_m})/(R_{m,i_1}+R_{m,i_2}+R_{m,i_3}).
\end{equation}
$R_{m,i_1}$, $R_{m,i_2}$, $R_{m,i_3}$ are same as the resistors in Fig.~\ref{fig:expona}, realized by transmission gates controlled by $V_{i_1}, V_{i_2}$ and $V_{i_3}$, similar to that in Fig.~\ref{fig:branch}.  
If any of the three variables $s_i$ in clause $C_m$ is satisfied, the corresponding $V_i$ turns off the respective transmission gate and cut off the current path from $V_{DD}$ to the capacitor. This circuit guarantees the continuous growth of $a_{m}$ since $V_{a_m}$ is charged by $V_{DD}$ till it reaches its upper bound $V_{DD}$ or any of the three variables in the clause is satisfied.

%This circuit provides $a_m$ with negative exponential growth rate, which may lead to a local minimum during the circuit diffusion process, it works well, however, in the problems which are not very hard and with small variable sizes. We will discuss more in the evaluation section via statistics as well as simulation results.

It is important to note that as the circuit in Fig.~\ref{fig:am-simple} does not realize the exponential growth specified in \eqref{dyn_a}, 
%thus may impact the effectiveness of avoiding being stuck in a local minimum during the gradient descent search process. 
it can indeed get captured into non-solution  attractors indefinitely for some very hard formulas. 
However, we have found that even for many hard problems, it works more efficiently than the op-amp based $a_m$ element (with the same threshold value) in finding solutions for smaller size problems (as long as they are solvable), and the dynamics would only rarely get stuck.
We will discuss this aspect more in the evaluation section via simulation results.

%The \av circuit introduced in Fig.~\ref{fig:am} does indeed help in many cases the CTDS to avoid being trapped in local minima and evolve towards global optimal solutions. We will also demonstrate this via simulation results in the next section. 
%design analysis above suggests that the limitation of $a_m$ has significant impact on the circuit capability to solve the problem. Both the theory formulas and the solver core simulation prove that the growing $V_{am}$'s enable signal fluctuations in the signal domain [0, 1]$^N$ and guarantee the sufficient magnitude (either positive or negative) of the signal dynamics $\frac{ds_i}{dt}$ to escape from a non-solution local trap domain. 
%Besides the ineffectiveness the simpler \av element brings in solving hard and large SAT problems, 

Similar to the op-amp based \av, in the simpler \av design in Fig.~\ref{fig:am-simple},  some $V_{a_m}$'s may reach $V_{DD}$ before the CTDS  converges to a solution.
%In this case, if the dynamics magnitude is smaller compared with the non-solution trap depth, the unchanged $V_{a_m}$'s in turn get the circuit trapped and signals fluctuated within the local trap domain and unable to escape from it.
%{\cg Zoltan: I don't like the above sentence but not sure what is a good way to rewrite it. Could you help?}
One way to alleviate this physical limitation is to increase the range of $V_{a_m}$. However, such an approach has its limitations in practical circuits (e.g., the limited voltage supply allowed).  This, in fact, is a fundamental limitation due to the NP hardness of 3-SAT. Nonetheless, it is possible to improve the $V_{a_m}$ driving capability  in the CTDS  and increase the size of the hard problems that can be solved with the same physical range of $V_{a_m}$.
%Note: the time-delayed design cannot gaurantee the power efficiency improvement within the same amount of power/energy. In the view of circuit side, this time-delayed design just decreases V_am of satisfied clauses, increasing the possibility of finding out global minimum.
Below, we discuss an alternative implementation of the simpler $a_m$ element to demonstrate that it is worthwhile to investigate different implementations of the \av.

\begin{figure}[!b]
	%	\vspace*{-2ex}
	\centering
	\includegraphics[width=0.4\textwidth]{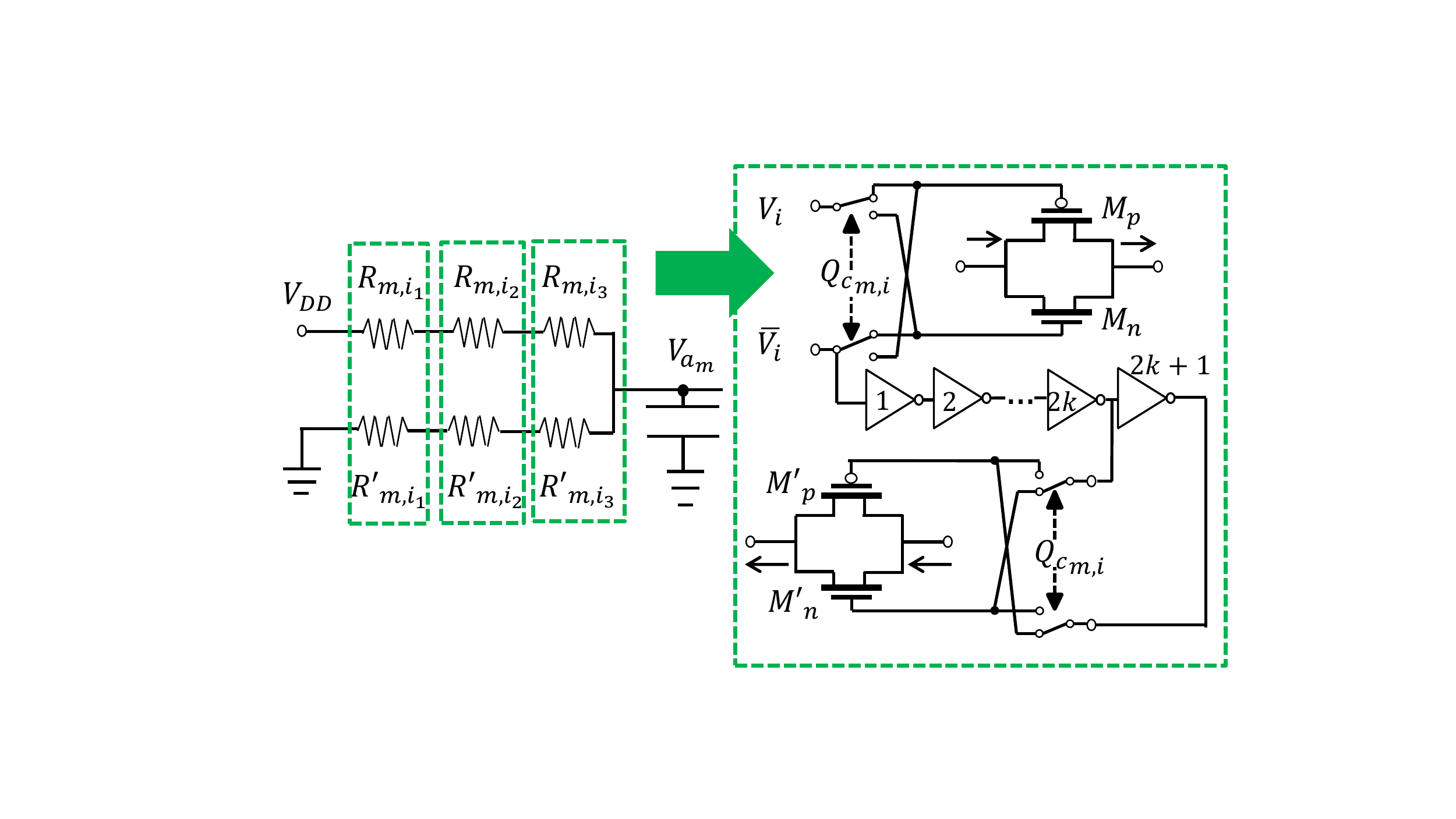}
	\caption{Circuit realization of the time-delayed simpler  $a_m$ growth. }
	\label{fig:newam}
	%	\vspace*{-2ex}
\end{figure}

%Here we consider the time-delayed form of $a_m$ given in \eqref{dyn_a_new}.
%Different from simply increasing the range, a modified $a_m$ variable dynamics equation using time-delays is proposed. 
Recall that the delay function $\delta_m(t)$ in \eqref{dyn_a_new} is to assist $a_m$ to keep relevant information from a limited range of the trajectory's past history instead of the entire history.  We consider combining the simpler $a_m$ element with this time-delayed form, and choose $\delta_m(t) = \delta$  (meaning that we are integrating over a fixed time window of length $\delta$).
The corresponding $a_m$ element is shown in Fig.~\ref{fig:newam}. Capacitor $C$ is charged to $V_{DD}$ through three tunable resistive elements and discharged to $G\!N\!D$ through the other three resistive elements. The first order differential equation of $V_{a_m}$ can be written as
\begin{equation}
\label{newacurrent}
C\frac{dV_{a_m}}{dt} =\frac{V_{DD}-V_{a_m}}{R_{m,i_1}+R_{m,i_2}+R_{m,i_3}}+\frac{-V_{a_m}}{R_{m,i_1}'+R_{m,i_2}'+R_{m,i_3}'}
\end{equation}
The six resistive elements are implemented by transmission gates similar to those for $R_{m,i_2}$ in Fig.~\ref{fig:branch}.  Specifically, $R_{m,i_1}', R_{m,i_2}', R_{m,i_3}'$ are controlled by $\textbf{s}$'s  (representing all $s_i$'s) earlier values, i.e., $\textbf{s}(t-\delta_m)$. A chain of an odd number of analog inverters is used to realize the delay $\delta$ as shown at the right of Fig.~\ref{fig:newam}. If signals $\textbf{s}(t)$ reach their targets at time $t$,  the path from $V_{DD}$ to capacitor $C$ is cut off, while the discharge path is still conducting current. $V_{a_m}$ keeps decreasing until the discharge path is cut off after $\delta$. Hence the circuit properly implements the time-delayed simpler $a_m$ growth function.  
%{\cg >>XY: I think below sentence should be added in background, because we have not implemented a delay with step by step amplification.<< 
%A constant delay $\delta_m$ can be selected by starting with a small value and doubling it every time the dynamics is stuck or hits an upper threshold - this only requires a few iterations. Other variable delay functions will be considered in future work.}
% This time-delayed $a_m$ variable array thus achieves the  $a_m$ variable dynamics Equ.~(\ref{dyn_a_new}) with constant delay.

%The simulation result Fig.~\ref{fig:amwave} shows the original and new trajectories of $V_{am}$'s respectively. As the theory predicts, the satisfied clauses' $a_m$ voltages decrease after they reach their peak magnitude and lowers down their impact on the associated signal dynamics.

%\begin{figure}[!t]
%	\centering
%	\includegraphics[width=0.5\textwidth]{amtrajectories}
%	\caption{Improved one array for $a_m$ variable }
%	\label{fig:amwave}
%\end{figure}

%\input{evaluation}
% !TEX root = main.tex
\section{Evaluation}
\label{sec:evaluation}

\begin{figure}[!t]
	%	\vspace*{-2ex}
	\centering
	\includegraphics[width=0.45\textwidth]{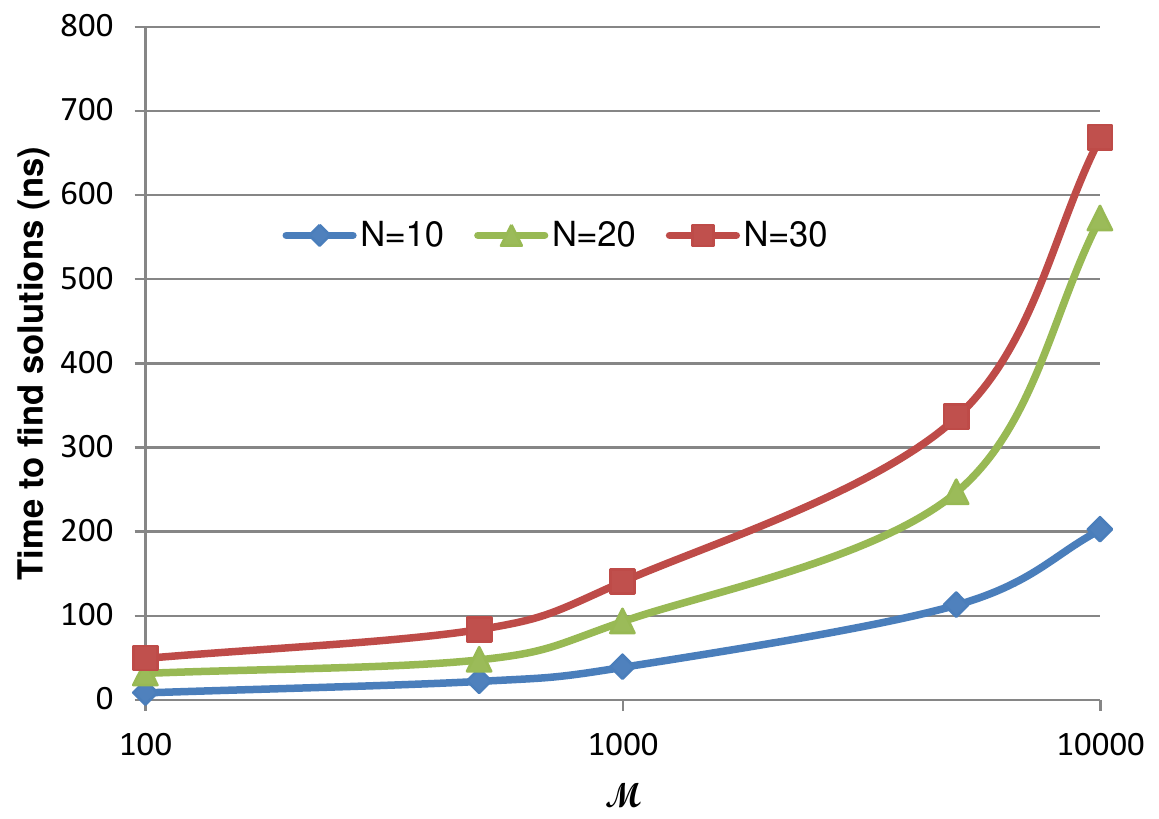}
	\caption{ SPICE simulation run times to find a solution with various \mssat circuit sizes. $N$ is the number of variables in a problem instance. The number of clauses in each problem instance is $4.25 \times N$ (corresponding to hard problems).}
	\label{fig:parasitics}
	%	\vspace*{-2ex}
\end{figure}

{\hu In this section, we present our evaluation study of \mssat. We first describe the basic functional validation and then discuss the robustness of \mssat against device variations. We finally compare the performance of \mssat with {\zt a state-of-the-art digital solver}.
	
	\subsection{Functional Validation }
	
	We have built our proposed analog SAT solver, \mssat, at the transistor level in HSPICE based on the PTM 32nm CMOS model~\cite{ptm}.  All the circuit components use $V_{DD}=1V$. 
	To achieve sufficient driving capability,  the minimum transistor size is set to  $W = 1\mu m$, $L = 40nm$  while actual transistor sizes are selected  according to their specific roles. For logic gates, the transistor sizes are chosen to ensure equal pull-up and pull-down strength. 
	For the Branch block in Fig.~\ref{fig:branch}, the relative W/L values of transistor $M_{p1}$ and $M_{n1}$ (i.e., the size of $R_{a_m}$) with respect to the W/L values of the other transistors (i.e., the sizes of $R_{m,i}$, $R_{m,i_2}$ and $R_{m,i_3}$) determine the contribution of $a_m$ to $I_{m,i}$. Thus,  tradeoff between the size of $R_{a_m}$ and the impact of $a_m$ should be carefully considered. In our implementation, since $R_{m,i}$ (dependent on the size of $M_{p2}$ and $M_{n2}$) is mainly used for proper current flow at the beginning, it should not dominate the current flow as $R_{a_m}$ starts  affecting $I_{m,i}$. Hence we chose the transistor sizes such that they result in the ratio  of $R_{a_m}$ to $R_{m,i}$, $R_{m,i_2}$ and $R_{m,i_3}$ being 64, 4 and 4, respectively. The sizes of the transistors in other circuits  are determined in a similar fashion.  Note that the transistor sizes shown above are just a lower bound for the technology model that we are using. The absolute values of the transistor sizes are not critical (the equations of the solver are adimensional), and other transistor sizes should also work as long as their relative sizes are close to the ones that we have shown.
	
	\begin{figure}[!t]
		\centering
		\vspace{-2ex}
		\subfigure[With $\epsilon_1e^{qt}a_m$ growth.]
		{\includegraphics[width=0.41\textwidth]{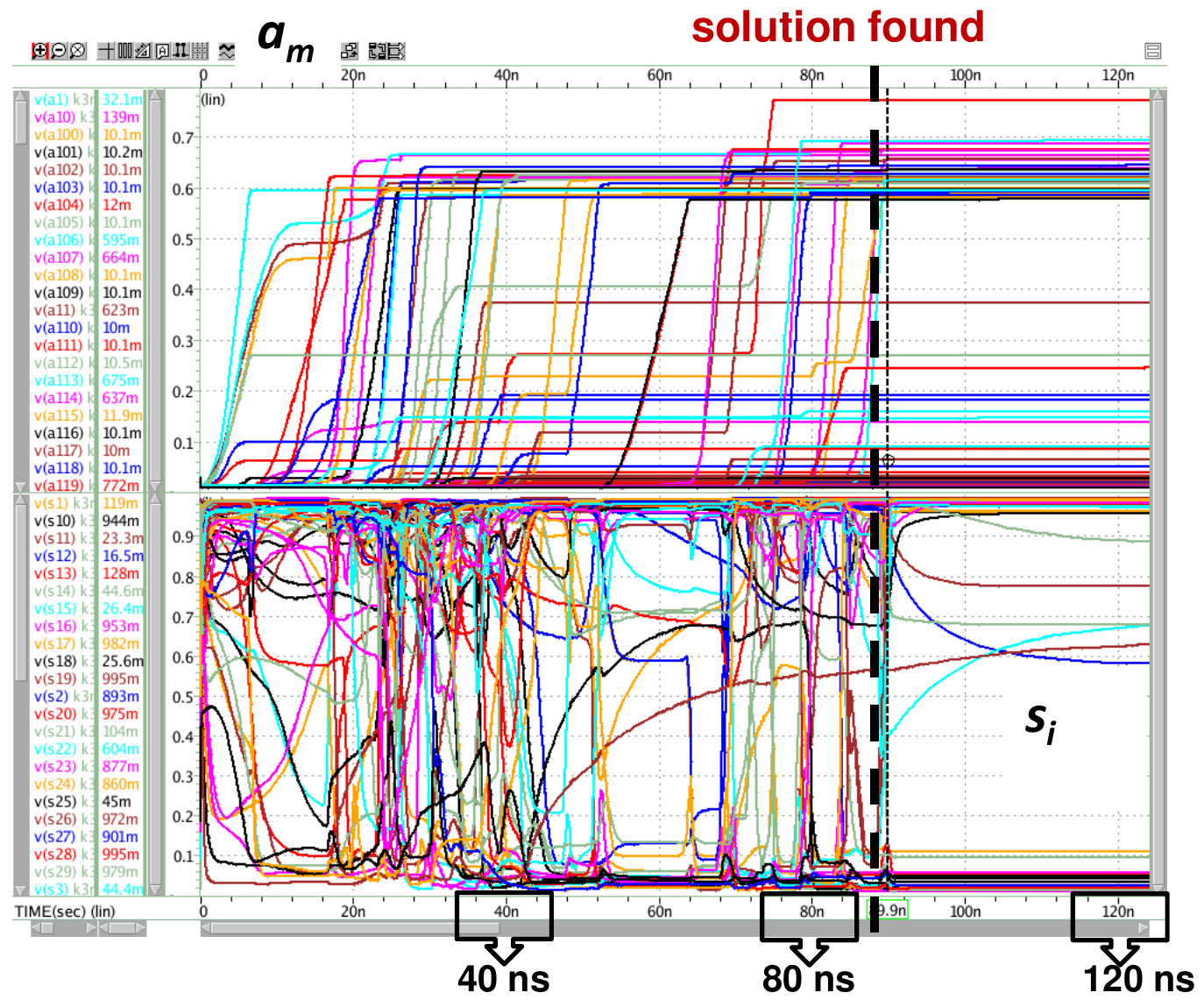}}  \vspace{-2ex}
		\subfigure[With $1-\epsilon_2e^{-qt}a_m$ growth.]
		{\includegraphics[width=0.41\textwidth]{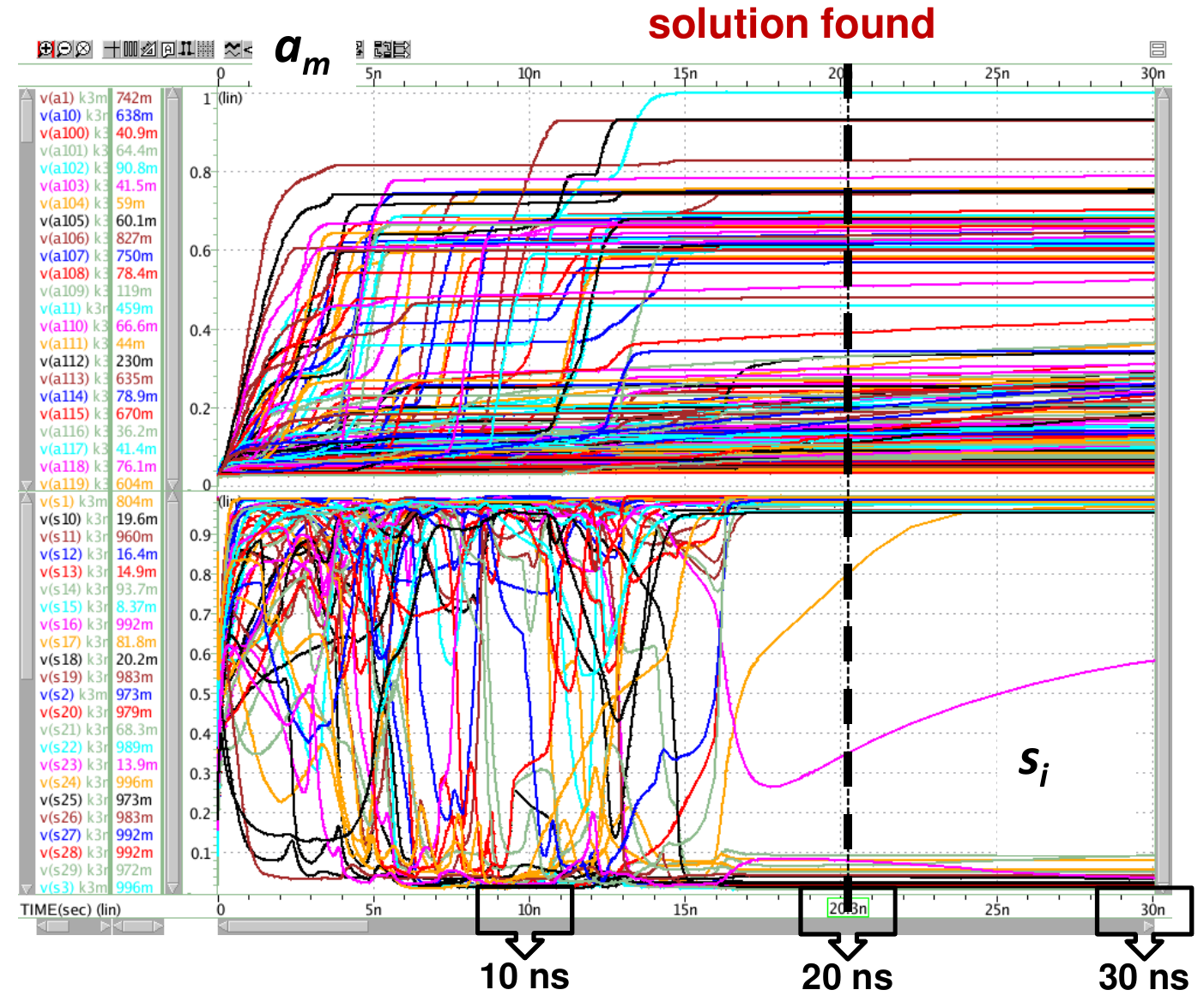}}
		\vspace{-2ex}
		\subfigure[With time delayed $a_m$ growth.]
		{\includegraphics[width=0.41\textwidth]{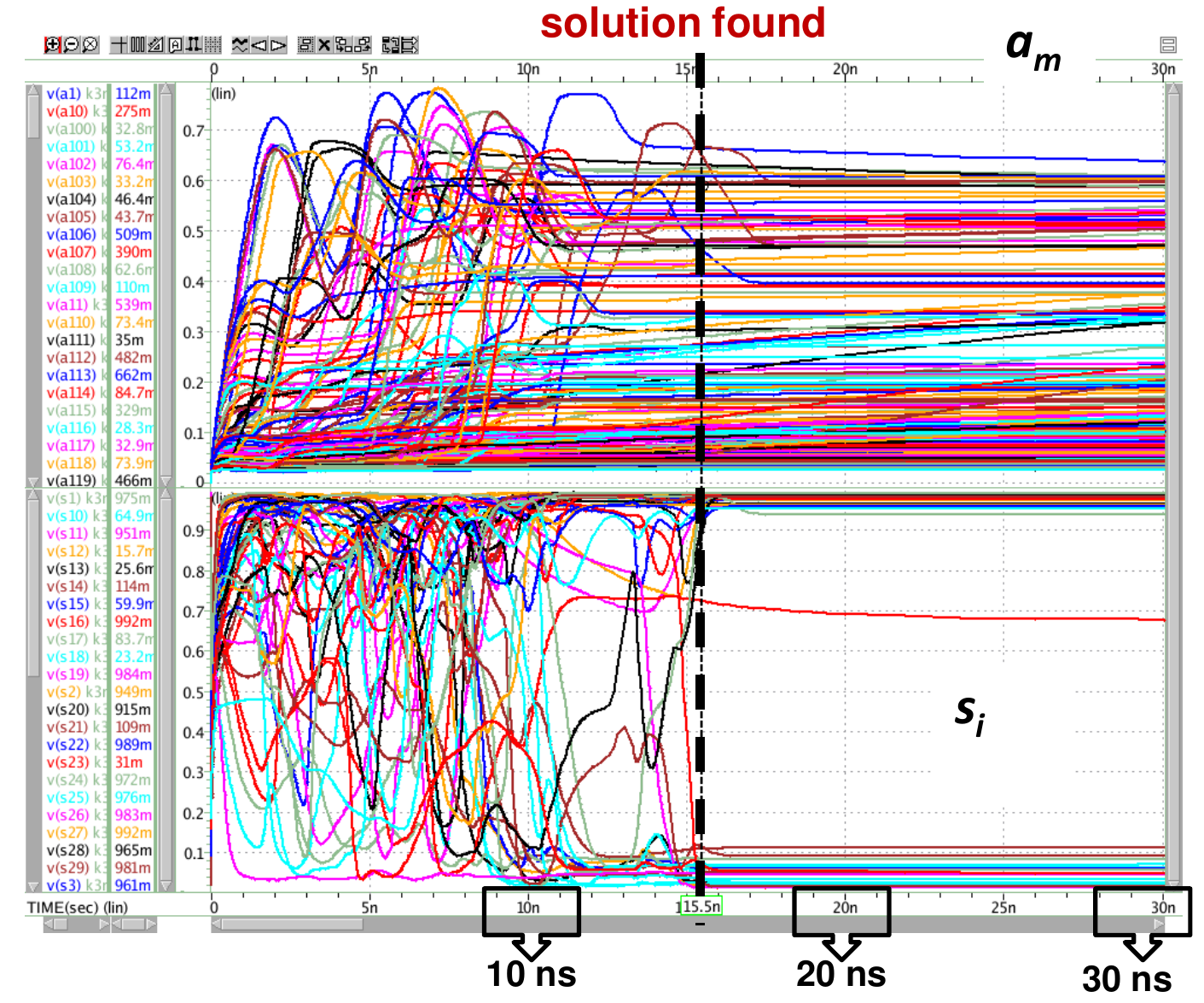}} 
		\caption{Waveforms of signals representing $s_i(t)$ and $a_m(t)$  for a 3-SAT problem with 212 clauses and 50 signal variables. The problems/formulas have a constraint density $\alpha$=$M/N$=4.25, and are considered to be hard problems.}
		%Left: \mssat with op-amp based \av circuit; right: \mssat with simpler $a_m$ implementation. exponential growth and negative exponential growth can be observed respectively.\
		
		\label{fig:testwave}
	\end{figure}
	
	To demonstrate that \mssat indeed behaves as specified by the CTDS dynamics in \eqref{dyn_s} and \eqref{dyn_a}, we examine the waveforms of signals $s_i$ and $a_m$.
	Fig.~\ref{fig:testwave} shows three sets of $s_i$ and $a_m$ waveforms from a 3-SAT problem instance having 50 variables and 212 clauses: Fig.~\ref{fig:testwave}(a) for the op-amp based $a_m$ implementation (realizing the  ($\epsilon_1 e^{q t}$)-type $a_m$ growth),  Fig.~\ref{fig:testwave}(b) for the simpler $a_m$ implementation (realizing the  ($1-\epsilon_2 e^{-q t}$)-type $a_m$ growth), and Fig.~\ref{fig:testwave}(c) for the time-delayed simpler $a_m$ implementation. For all three designs, \mssat successfully finds a solution after a certain time as indicated by the vertical dashed lines.  Note that \mssat determines whether a solution is found  via the \dv. As can be seen from the $s_i$ trajectories, the $s_i$ signals stabilize (i.e., converge) after a solution is found. Comparing the $a_m$ trajectories in the three different designs, one can see that the $a_m$'s grow most rapidly in the op-amp based design due to the exponential growth function while some of the $a_m$'s (the ones corresponding to the satisfied clauses) in the time-delayed implementation decrease after they reach their peak magnitude, just as predicted by \eqref{dyn_a_new}.

	\subsection{Scaling Considerations}}
\label{sec:parasitics}
Besides functionality, the impact of interconnect parasitics on the circuit is another important consideration towards practical and modular designs of the solver. As the circuit size increases (i.e., O(\cm\cn)), for each variable array element in the \sd, the total parasitic capacitance from the Branch blocks (Fig. \ref{fig:xbranch}) increases linearly with the number of Branch blocks \cm, namely the maximum number of clauses that the solver can handle. Given a problem instance, if variable $x$ is involved in $y$ clauses ($y<\cm$), then $y$ Branch blocks associated with $x$ are active, while all other Branch blocks are turned off. However, all the Branch blocks contribute parasitic capacitance to the dynamical evolution of the variable $x$. To investigate the impact of parasitic capacitance, we have conducted a number of simulations of the solver circuit with various number of Branch blocks (i.e., \cm) in the \sd, i.e., 100, 500, 1000, 5000, 10000 Branch blocks for each variable array element. We used the circuits to solve various problem instances with 10, 20, 30 variables, and evaluated the time to find a solution. Simulation results shown in Fig. \ref{fig:parasitics} demonstrate that as the solver circuit becomes larger, the solver still functions correctly, but takes longer time to find solutions due to larger parasitic capacitance.

Another issue due to interconnect scaling is the capacitance value associated with the \av elements. As the parasitic capacitance associated with variable signals increases with the number of Branch blocks, the dynamic evolution of the variable signals becomes slower due to the RC charging rule. As a consequence, the AVC element, whose internal capacitance (i.e., contributed by the two capacitors in the AVC element) is much smaller than the parasitic capacitance associated with the variable signals,  will charge $V_{a_m}$ quickly, and reach the $V_{a_m}$ upper bound before the variable signals find the solution. This trend could make the \av element less effective and thus lead to the solver not able to find a solution. Therefore, it is critical to increase the values of the capacitors in the \av elements as the circuit size increases. A basic approach is to choose the capacitance value  such that the $V_{a_m}$'s  RC constant is comparable or smaller than the variable signal RC constant. As the parasitic capacitance of the \sd increases linearly with the number of Branch blocks, the capacitance in the \av element should also scale proportionally. 
%CTDS theory. 
%It is important to note that the time required to find the solution and the evolving behavior of $s_i$ signals heavily depend on the initial conditions of the signals $\textbf{s}(t)$ as well as the problem size. Moreover, the limited physical constraint further enhances the dependency of the circuit capability to solve problems on initial conditions, that is, a set of initial conditions may find the solution before the circuit hits the supply voltage limit while another set of initial conditions may not. However, exploring the impact of initial conditions is beyond the scope of this paper, and we will focus on the comparison between the two \mssat solvers in terms of the capability to solve 3-SAT problems.

%\begin{figure*}[!t]
%\centering
%\includegraphics[width=0.95\textwidth]{k3m212n50wave2}
%\caption{Waveforms of signals representing $s_i(t)$ and $a_m(t)$  for a 3-SAT problem with 212 clauses and 50 signal variables. The problems/formulas have a constraint density $\alpha$=$M/N$=4.25, and are considered to be hard} problems. 
%Left: \mssat with op-amp based \av circuit; right: \mssat with simpler $a_m$ implementation. exponential growth and negative exponential growth can be observed respectively.\

%\label{fig:testwave}
%\end{figure*}

{\hu \subsection{Device Variation Study}}

\begin{figure}[!t]
	\centering
	\includegraphics[width=0.45\textwidth]{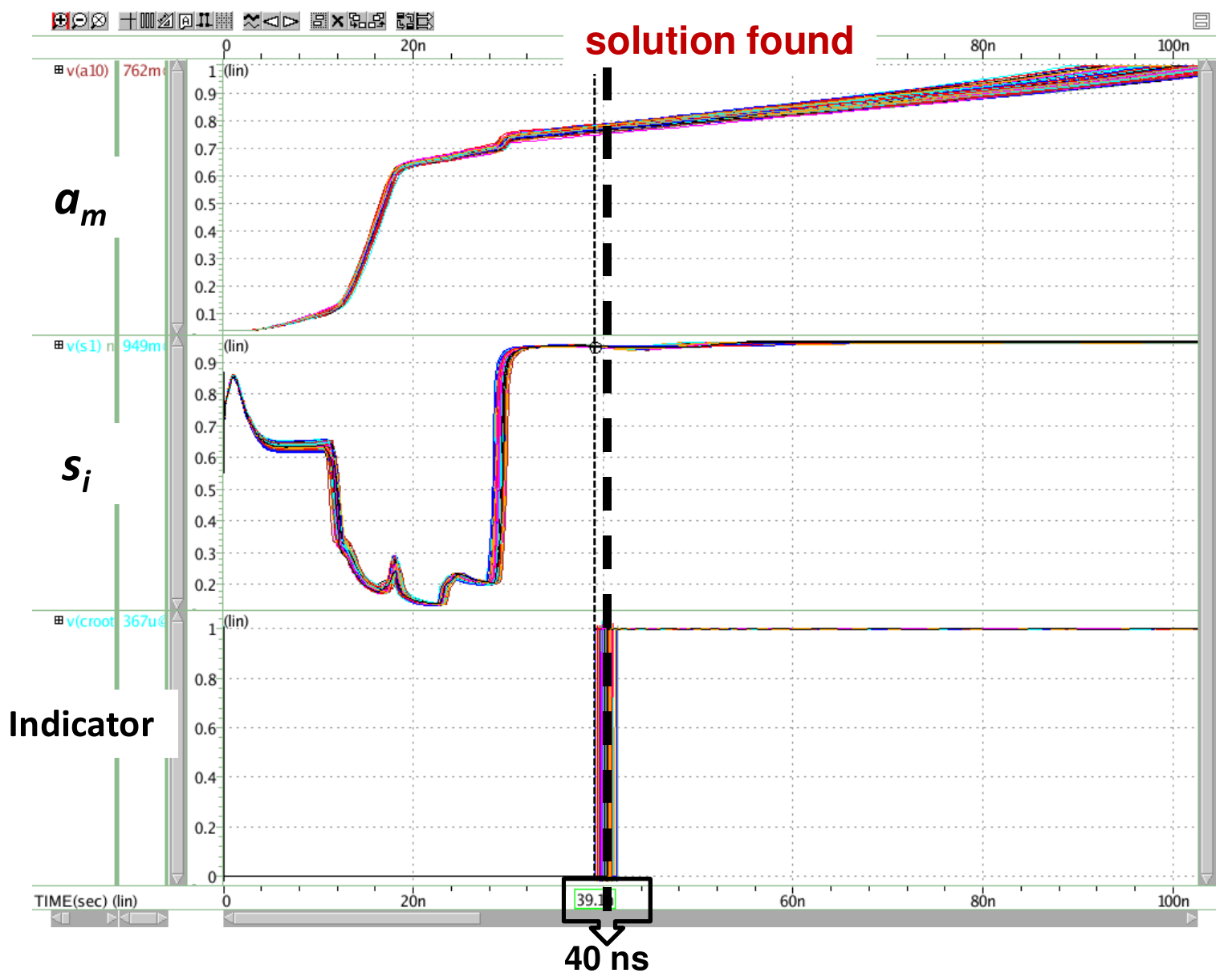} % \vspace*{1ex}
	% \vspace*{-1ex}
	\caption{ Waveforms of signals representing a single $a_m(t)$ and $s_i(t)$ for a 3-SAT problem instance (op-amp based design) with 42 clauses and 10 signal variables under Monte Carlo simulations. The third row represents the output of \dv, which turns to $V_{DD}$ once the solution is found.}
	%Left: \mssat with op-amp based \av circuit; right: \mssat with simpler $a_m$ implementation. exponential growth and negative exponential growth can be observed respectively.\
	
	\label{fig:MCsim}
\end{figure}

After validating that \mssat indeed can solve SAT problems {\zt correctly}, we further investigate the robustness of \mssat against device variations. Typical analog circuits can be rather sensitive to device variations if not designed well. However, \mssat has two unique advantages in this aspect. First, the circuit itself does not rely on device matching. Secondly, the CTDS theory has been shown in theory to be robust against {\zt noise~\cite{sumi2014robust}}.  To demonstrate the robustness of our proposed \mssat system, we have conducted Monte Carlo simulations with respect to transistor size variations for randomly chosen 3-SAT problems. Specifically, we let the transistor widths follow a Gaussian distribution with standard deviation ($\Delta W/W$) of $0.05\mu m/\sqrt{W\times L}$  for all transistor widths,  which is an acceptable variance distribution for the 32nm technology node \cite{nikandish2007performance}.  In other words, the solver circuit is simulated with the Monte Carlo method considering 5\% transistor width variations. For each problem, 100 Monte Carlo runs were performed.  Fig.~\ref{fig:MCsim} shows the waveforms of one $a_m(t)$ signal and one $s_i(t)$ signal plus the output of \dv for one problem instance for 100 Monte Carlo simulations. As can be seen from the signal trajectories, the signals evolve consistently in the Monte Carlo simulations, and the results demonstrate the robustness of the circuit. Moreover, since analog circuits generally use mature technology nodes, (e.g. 180nm, 90nm), we in fact validated our design in a relatively aggressive way. The circuit is expected to perform much better under mature technologies, whose variations would be much smaller than 5\%.

%Readers may notice that the time scale of the waveforms in Fig.~\ref{fig:testwave}(a) is different from that of the other two sets of waveforms. For this particular problem instance,  the op-amp based implementation takes  much longer time to find a solution than the other two designs. 
To get a better comparison between the different designs,  we {\hu performed Monte Carlo simulations on both the op-amp based and the simpler $a_m$ based \mssat. The two designs are used} to solve  2000 randomly generated, hard ($\alpha=4.25$) 3-SAT problems containing 1000 instances of a small problem size (N=10) and 1000 instances {\zt of a larger} problem size (N=50). It has been verified that all these 2000 instances are solvable. With the same fixed supply voltage and initial conditions,   \mssat with the op-amp based $a_m$ design solves 91.1.\% of the N=10 instances and 58.2\% of the N=50 instances, respectively, while \mssat with the simpler $a_m$ design solves 86.9\% of the N=10 instances and 46.5\% of the N=50 instances, respectively.  (Note that \mssat did not solve all the problems because of the physical {\zt voltage limit we imposed}.\footnote{This is not a limitation of the CTDS theory, but rather the supply voltage bound set in our design. In fact, our software implementation of CTDS is able to solve all the problems. Relaxing the voltage bound will help solve more problem instances, which is left for future work.})   These results indicate  that, as expected, within the same physical constraints, \mssat based on the exponential growth $a_m$  is more effective than \mssat with the $(1-\epsilon_2 e^{- q t})$-type  $a_m$ growth. {\hu  Note that an exponential growth $a_m$ circuit implemented with an op-amp  does consume larger area and energy, while the simpler $a_m$ circuit trades off area and energy with solver capability.}

\medskip

%The simulation result Fig.~\ref{fig:amwave} shows the trajectories of original $a_m$ and new $a_m$ implementing alternative design in Sec.~\ref{sec:improvedesign} respectively. As the theory predicts, the satisfied clauses' $a_m$ voltages decrease after they reach their peak magnitude and lowers down their impact on the associated signal dynamics.

{\hu \subsection{Performance Comparisons}}

To further investigate the effectiveness of \mssat, we compare the simpler $a_m$ based \mssat design with (i) a software program that solves the system \eqref{dyn_s}-\eqref{dyn_a} using an adaptive Runge-Kutta, fifth-order Cash-Karp method and (ii) the software MiniSat solver~\cite{een2003extensible}. The software programs are running on the same digital computer. %with Intel Core i7-4700 running at 2.4 GHz. 
We randomly generated  5000 hard ($\alpha=4.25$) 3-SAT  problems  that contain 1000 instances for each problem size  of N=10, 20, 30, 40, 50. The same initial conditions  are applied whenever appropriate. 
%We plot the fraction of problems $p(t)$ not solved by continuous time $t$ in Fig.~\ref{fig:stats} and highlight the time required for each solvers to reach $p$=0.1 unsolved fraction. 
Table~\ref{table:stats} summarizes the average time needed to find solutions for each problem size. The \mssat column reports the analog/physical times taken by \mssat.
The CTDS and MiniSat columns report the CPU times of the two software implementations, respectively. (To be fair, only the times taken by the solved problems for all three methods are included.) Observe that the times in the CTDS column increase nearly exponentially as the problem size increases. This is  natural, since the numerical integration happens on a digital Turing machine, and in order to ensure the pre-set accuracy of computing the chaotic trajectory the Runge-Kutta algorithm has to do a very large number of window-refining discretization steps.
%We could see that for size N=10, the time consumed to solve 90\% of the 1000 instances for analog circuit solver is at $\sim$10 ns, while for software solver of CTDS the time is at $\sim$1 ms, and for Minisat solver the time is at $\sim$10$^2$ $\mu$s. 
As seen from the data in Table~\ref{table:stats}, \mssat demonstrates average speedup factors of $\sim$10$^5$ to $\sim$10$^6$ and $\sim$10$^4$ over software CTDS and MiniSat, respectively.  
%This result suggests that the CTDS theory can be a good match for analog circuits  that diffuses signals following its energy optimization. {\cg $\leftarrow$ I still do not get this formulation about diffusion of signals (ZT). Can you please change it?} 

\mssat is also very competitive compared with existing hardware based approaches. For example, a recent work~\cite{thong2013fpga} reported a CPU+FPGA based MiniSat solver achieving $\sim$4X performance improvement  over CPU based MiniSat. Since ASIC implementations typically achieves a maximum of 10X performance improvement over their FPGA counterparts~\cite{kuon2007measuring}, compared with a projected ASIC version of the FPGA design in~\cite{thong2013fpga}, \mssat would still result in $\sim$600X or higher speedup. 
We do not directly compare with the custom digital IC in \cite{gulati2010accelerating}
since our simulation-based system cannot solve the large size problems considered in \cite{gulati2010accelerating}. (Note that the total solving times reported in \cite{gulati2010accelerating} are extrapolated instead of directly obtained from simulation.) It is reported in \cite{gulati2010accelerating} that an average speedup of $\sim$10$^3$X over CPU based MiniSat is obtained. As contrast, \mssat achieves $\sim$10$^4$X speedup over CPU based MiniSat.
T%he average power consumption of \mssat for the size N=50 problems is 130 $mW$ which is mostly due to the static power dissipation by the analog inverters and  {\texttt{NAND}} gates.

\begin{table}[!b]
	%\vspace*{-4ex}
	\caption{Performance comparison of AC-SAT, software CTDS and MiniSat}
	\label{table:stats}
	\centering
	\begin{minipage}{3.4in}
		\centering
		\begin{tabular}{c |c| c c c}
			\hline\hline
			\multicolumn{2}{c|}{SAT solver}  & AC-SAT & CTDS & MiniSat\\
			\hline
			\multicolumn{2}{c|}{Platform}  & \begin{tabular}[!t]{@{}c@{}}ASIC \\ 32nm CMOS\end{tabular}
			& \begin{tabular}[!t]{@{}c@{}}Intel Core\\ i7-4700 \\ @2.4 GHz\end{tabular}
			& \begin{tabular}[!t]{@{}c@{}}Intel Core \\i7-4700 \\ @ 2.4GHz\end{tabular} \\
			\hline
			\multirow{5}{9ex}{Average time for each size N (s)} 
			& N=10 & 4$\times$10$^{-9}$ & 4.40$\times$10$^{-4}$ & 2.3$\times$10$^{-4}$ \\
			& N=20 & 7$\times$10$^{-9}$ & 3.91$\times$10$^{-3}$ & 2.4$\times$10$^{-4}$ \\
			& N=30 & 10$^{-8}$ & 1.62$\times$10$^{-2}$ & 2.8$\times$10$^{-4}$ \\
			& N=40 & 1.2$\times$10$^{-8}$ & 5.22$\times$10$^{-2}$ & 3.1$\times$10$^{-4}$ \\
			& N=50 & 1.4$\times$10$^{-8}$ & 1.13$\times$10$^{-1}$ & 3.7$\times$10$^{-4}$ \\
			\hline\hline
		\end{tabular}
		\\
		%	\begin{flushleft}
		%		$*$: assuming total area (including wirings) is twice the area of transistors\\
		%		$\star$: high level of $CLK$.\\
		%		Note: In SPICE simulation, a 100MHz signal is applied to $CLK$. All simulations are done with a FO4 inverter as load.
		%	\end{flushleft}
	\end{minipage}
	%	\vspace*{-3ex}
\end{table}

Readers may be concerned with the complexity of the analog hardware design as well as other issues such as noise. It is important to note that the analog solver core is modular  and  consists of arrays with the same  topology. %(Fig.~\ref{fig:xbranch},~\ref{fig:branch},~\ref{fig:am}), supply rails ($V_{DD}$, $-V_{SS}$) and inputs (problem matrix \textbf{C}), which enables the scalability as problem size increases. 
Furthermore, the CTDS theory has been shown to be robust against  noise~\cite{sumi2014robust}. \mssat is programmable, which means that different problem instances can be programmed or mapped to the \mssat circuit. \mssat is also modular, implying that (i) it can be more easily extended to construct a larger solver, and (ii) multiple \mssat components can be used to solve the same problem instance by providing different initial conditions, hence allowing larger space to be searched simultaneously.

The current implementation of \mssat,  however, does have some limitations.  In particular, while the modular structure allows  possible expansion to solve problems with larger numbers of variables and clauses, it can only address problems with clauses that have no more than the given number of $k$ literals ($k=3$ here).   One way to solve such problems is to use the host processor to convert $k$-SAT problems (where $k>3$) to 3-SAT problems (which can be done in polynomial time \cite{GareyJohnson79}). How to directly tackle such challenges in hardware is left for future work.

%can handle any SAT problem in which the clauses have at most $k$ literals, at most $N$ variables, and at most $M$ clauses (for which the circuit was designed),  it cannot handle problems where any of these parameters are larger, i.e., with more than $N$ variables, or with more than $M$ clauses, or when a clause has more than $k$ literals.} We plan to address these challenges in our future work. 

%In spite of the fact that there always exists a problem whose size is larger than the capacity of the solver core, efforts can be made on theory side to decompose such a problem and feed smaller subproblems into the solver core. %That would be another story.

%\begin{figure}[!t]
%	\centering
%	\includegraphics[width=0.5\textwidth]{trajectories}
%	\caption{$a_m$ trajectories of top: simpler $a_m$ implementation and bottom: time-delayed $a_m$ implementation on the same problem. Decrease in $a_m$ is observed.}
%	\label{fig:amwave}
%\end{figure}

%\begin{figure*}[!t]
%	\centering
%	\includegraphics[width=\textwidth]{stats}
%	\caption{The fraction of problems $p(t)$ not yet solved by continuous time $t$ for 3-SAT at $\alpha$=4.25 for N=10, 20, 30, 40, 50. HSPICE simulations are done over 1000 instances for each N. For each instance the signal variables are started from the same initial condition. The statistics are collected for analog CTDS circuit solver (6-Core Intel Xeon X5670 Cpu @2.93 GHz), software CTDS solver and Minisat solver (on Intel Core i7-4700 CPU @ 2.4GHz core) from left to right.}
%	\label{fig:stats}
%\end{figure*}

%\input{conclusion}
% !TEX root = main.tex
\section{Conclusions} \label{sec:discussions}

We presented a proof-of-principle analog system, \mssat,  based on the CTDS in \cite{NatPhys_ET11} to solve 3-SAT problems. The design can be readily extended to general $k$-SAT problems. \mssat is modular, programmable and can be used as a SAT solver co-processor. In this implementation the circuit size grows  polynomially (O(N$^2$)) as the problem size increases. Three different design alternatives were proposed and verified for implementing the auxiliary variable dynamics required by the CTDS. Detailed SPICE simulation results show that {\hu \mssat can indeed solve SAT problems efficiently and can tolerate well device variations.}  Compared with other SAT solvers, \mssat can achieve $\sim$10$^4$X speedup over MiniSat running on a state-of-the-art digital processor, and can offer over 600X speedup over projected digital ASIC implementation of MiniSat. %{\cg Furthermore, though \mssat experiments have shown that the limitation of the circuit is the physical condition, i.e. supply voltage, not the theory. } 

Regarding the practical use of a hardware solver, we note that there are instances in the SAT contests that take a very long time (e.g. days or even months) to solve. The reason for the long (and exponentially growing) running time is due not only to the size of the problems, but also to their hardness. It has been {\cg demonstrated} that when the constraint density ($M/N$) of a problem instance is between 4 and 5 (for 3-SAT), the problem can be very hard and take exponentially growing time for current software solvers to find a solution. Our work, together with its theoretical basis, however, provides a means to trade time for energy in order to speed up computations. With the circuit-friendly theory and proof-of-principle hardware implementation, we can solve hard SAT problems much faster than with software solvers on digital machines, however, at the expense of other resources such as energy (voltage and or/current values). Such tradeoffs are desirable for certain time-sensitive problems.

The CTDS equations (especially the dynamics for the auxiliary variables) and their analog implementations are not unique. It is quite possible that better forms and implementations exist. The fact that our proof-of-principle circuit implementations significantly outperform state-of-the-art solvers on digital computers are an indication that analog hardware SAT solvers  have a great potential as application-specific processors for discrete optimization. As future work, we will further investigate alternative implementations of the auxiliary variable dynamics as well as methods to handle problem instances that do not fit on a given hardware implementation, e.g., through problem decomposition. Moreover, we will explore other methods that can, in principle, solve SAT problems even more efficiently, e.g., by combining clause learning (handled by a digital processor) with our analog solver.
%We also plan to realize \mssat on hardware, first on a field-programmable analog array (FPAA) board and eventually on a custom chip.
%Specific challenges associated with analog hardware  such as noise and fabrication variations will also be further investigated.

%uses only linearly many ${O(N)}$ parallel running branches, and thus, for the hardest problems on occasion the auxiliary variables will need to grow exponentially, at least until the dynamics is pulled out of local minima. In the current implementation, the circuit was based on auxiliary variables that have grown only polynomially fast to a preset saturation value, however, even this system could solve many of the hard 3-SAT problems without getting stuck, taking only on the order of tens of nanoseconds to find the solutions. 
%A full circuit implementation of the equations with on-demand exponential acceleration is currently under development. 

\vspace*{2ex}
\noindent
{\bf \large Acknowledgement:} 
The authors acknowledge useful discussions with S. Datta, 
A. Raychowdhury, M. Niemier and G. Cauwenberghs. {\cgb The authors also thank anonymous reviewers for helpful comments. }This project was supported in part by the National Science Foundation under grant numbers CCF-1644368 and 1640081, and the Nanoelectronics Research Corporation (NERC), a wholly-owned subsidiary of the Semiconductor Research Corporation (SRC), through Extremely Energy Efficient Collective Electronics (EXCEL), an SRC-NRI Nanoelectronics Research Initiative under Research Task ID 2698.004 (XSH,ZT). MER was funded by a European Commission Horizon 2020 Program Grant No. 668863-SyBil-AA and a Romanian CNCS-UEFISCDI Research Grant No. PN-III-P2-2.1-BG-2016-0252.

\bibliographystyle{IEEEtran}
\bibliography{bib-1}  % sigproc.bib is the name of the Bibliography in this case
%\input{bio}
% !TEX root = main.tex
\begin{IEEEbiography}[{\includegraphics[width=1in,height=1.25in,clip,keepaspectratio]{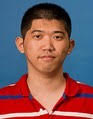}}]{Xunzhao Yin}
	(S'16) received his B.S. degree from in electronic engineering from Tsinghua University, China, in 2013. He is currently pursuing the Ph.D. degree in the department of Computer Science and Engineering, University of Notre Dame, Indiana, USA. He has been a research assistant in University of Notre Dame since 2013, and he is a member of Center for Low Energy System Technology (LEAST), where he is exploring the novel circuits and systems based on beyong-CMOS technologies. His research interests include hardware security, low-power circuit design and novel computing paradigms with both CMOS and emerging technologies. 
\end{IEEEbiography}
%\vspace*{-0.25in}
\begin{IEEEbiography}[{\includegraphics[width=1in,height=1.25in,clip,keepaspectratio]{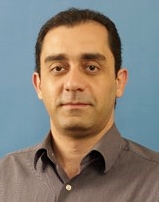}}]{Behnam Sedighi}
	received his Ph.D. degree in Electrical Engineering from Sharif University of Technology, Tehran, Iran, in 2008. From 2009 to 2011, he was a Senior Circuit Designer at IHP Microelectronics, Frankfurt (O), Germany, where he was working on analog/mixed-signal integrated
	circuits for optical communications. He then joined the center of energy efficient telecommunications and NICTA, University of Melbourne, Victoria, Australia, and subsequently, the center for low-energy systems technology at University of Notre Dame, IN, USA as a Research Associate. He is now with Qualcomm Inc. His research interests include broadband ICs, communication circuits and systems, data converters, and nanoelectronics.
\end{IEEEbiography}
%\vspace*{-0.25in}
\begin{IEEEbiography}[{\includegraphics[width=1in,height=1.25in,clip,keepaspectratio]{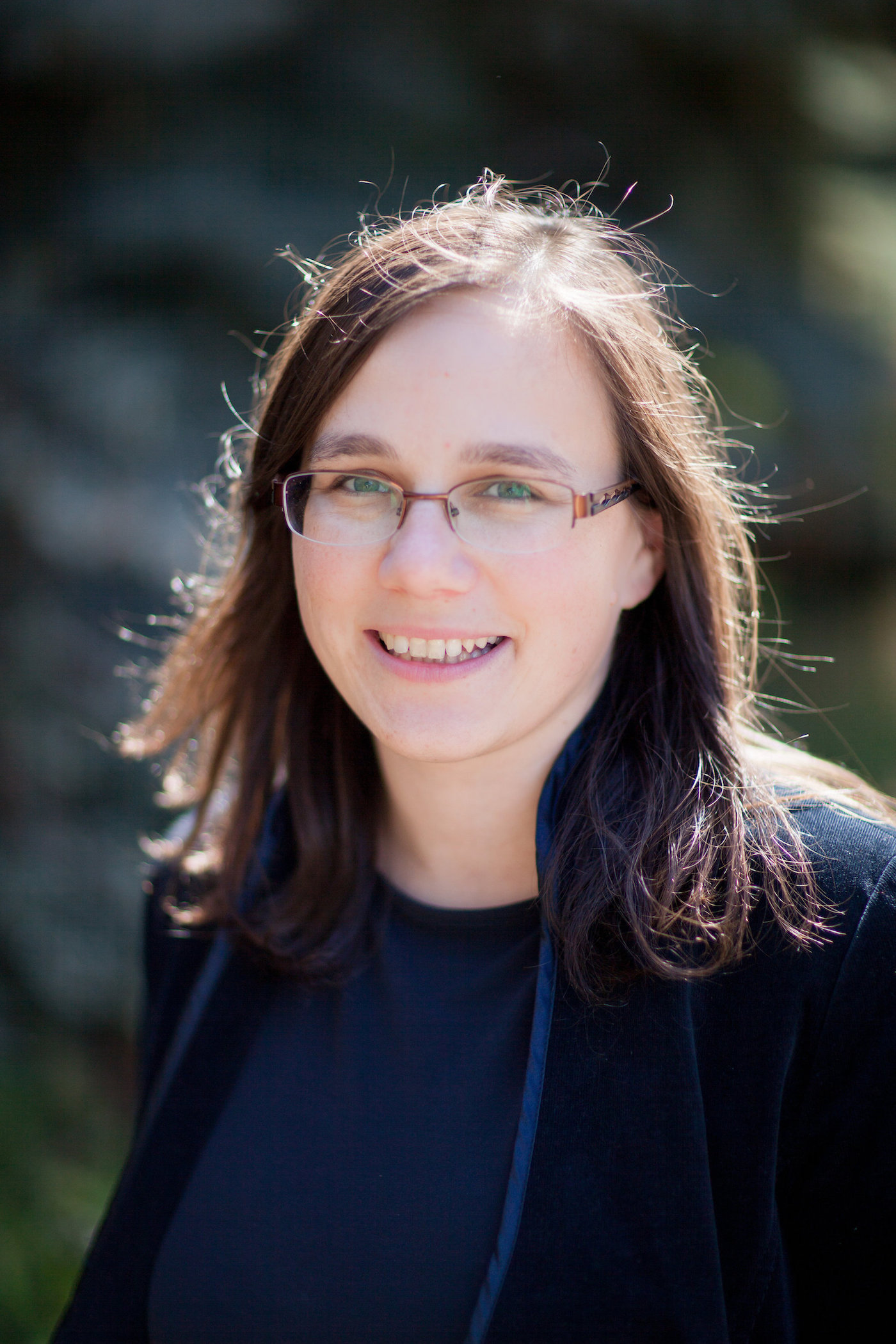}}]{Melinda Varga}
	Melinda Varga is a graduate student at Department of Physics at the University of Notre Dame, Indiana, where she works with Prof. Zolt\'an Toroczkai. She received her B.S. and M.S. degrees from Babes-Bolyai University, Romania. Her research interests include nonlinear dynamical systems and chaos theory, complex networks, neuroscience, statistical physics, evolutionary game theory. Her publications cover a wide range of topics. During her masters studies she received her institution's Research Performance Scholarship. She was one of the finalists for the GSNP Student Speaker Award at the APS March Meeting in 2016.
\end{IEEEbiography}
%\vspace*{-0.45in}
\begin{IEEEbiography}[{\includegraphics[width=1in,height=1.25in,clip,keepaspectratio]{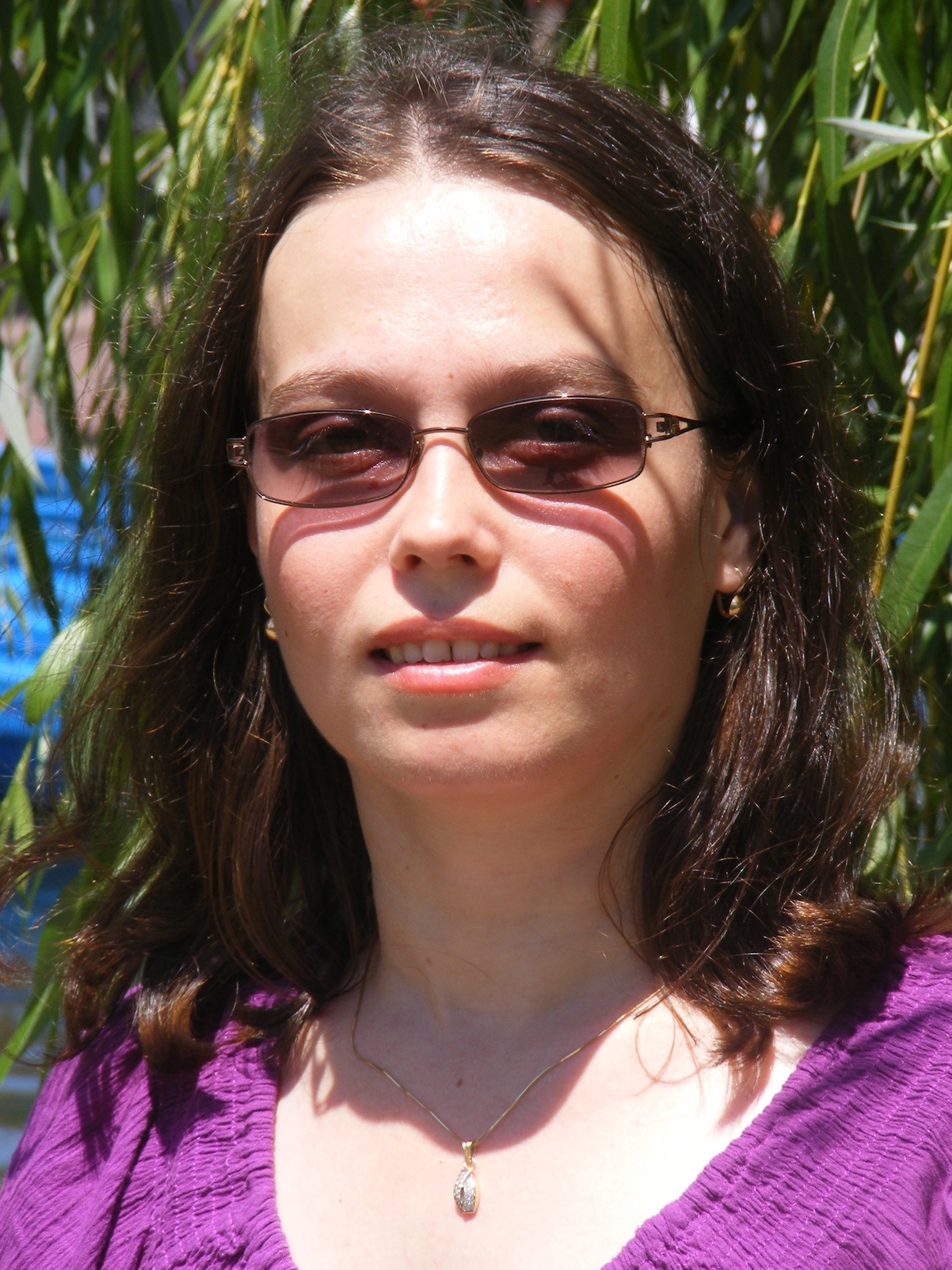}}]{Maria Ercsey-Ravasz}
	received her B.Sc. and M.Sc degree in Physics from the Babes-Bolyai University, Cluj-Napoca, Romania. In 2008 she obtained a PhD joint degree in Physics at Babes-Bolyai University and in Information Technology (Infobionics) at Peter Pazmany Catholic University, Budapest, Hungary. After a three year postdoctoral fellowship at  University of Notre Dame, iCeNSA, she returned to Romania with a Marie Curie Fellowship. She is currently researcher at the Babes-Bolyai University and the Romanian Institute of Science and Technology.  Her research interests include  network science with applications in different domains, such as neuroscience; analog computing; optimization problems and  nonlinear dynamics. She has published more than 30 ISI articles. In 2013 she got the UNESCO-L'Oreal National Fellowship "For Women in Science". In 2015 she received the Constantin Miculescu Award of the Romanian Academy of Sciences.
\end{IEEEbiography}
%\vspace*{-0.45in}
\begin{IEEEbiography}[{\includegraphics[width=1in,height=1.25in,clip,keepaspectratio]{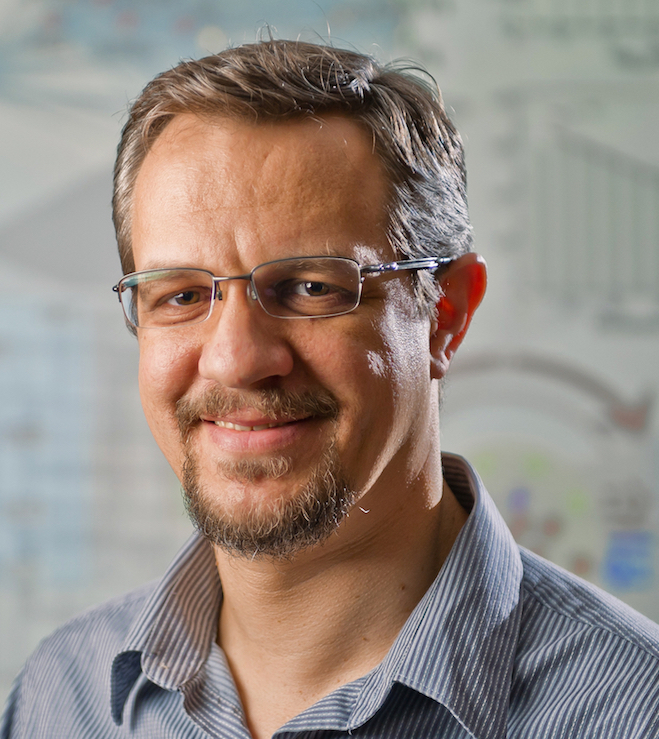}}]{Zoltan Toroczkai}
	received his Ph.D. degree (1997) in theoretical physics from Virginia Polytechnic Institute and State University, Virginia. He is a Professor in the Department of Physics  and a Concurrent Professor in the Department of Computer Science and Engineering at University of Notre Dame. He spent his postdoctoral years in the condensed matter
	physics group at University of Maryland at College Park, then as a Director Funded Fellow at Los Alamos National Laboratory (LANL). He then joined the complex systems group at LANL as a regular research staff member. In 2004, he became the Deputy Director of the Center for Nonlinear Studies at LANL until his joining of the department of physics at University of Notre Dame in 2006. His research interests lie in the areas of statistical physics, nonlinear dynamical systems and mathematical physics with topics including fluid flows, reaction kinetics, interface growth, population dynamics, epidemics, agent-based systems and game theory, complex networks, foundations of computing, and brain neuronal systems. He has authored and coauthored over 90 peer-reviewed publications on these topics. He has been on the editorial board at European Journal of Physics B, Scientific Reports, Chaos and as Associate Editor for Network Science.  He was elected APS Fellow in 2012 upon nomination by GSNP ''For his contributions to the understanding of the statistical physics of complex systems, and in particular for his discoveries pertaining to the structure and dynamics of complex networks''.
\end{IEEEbiography}
%\vspace*{-0.45in}
\begin{IEEEbiography}[{\includegraphics[width=1in,height=1.25in,clip,keepaspectratio]{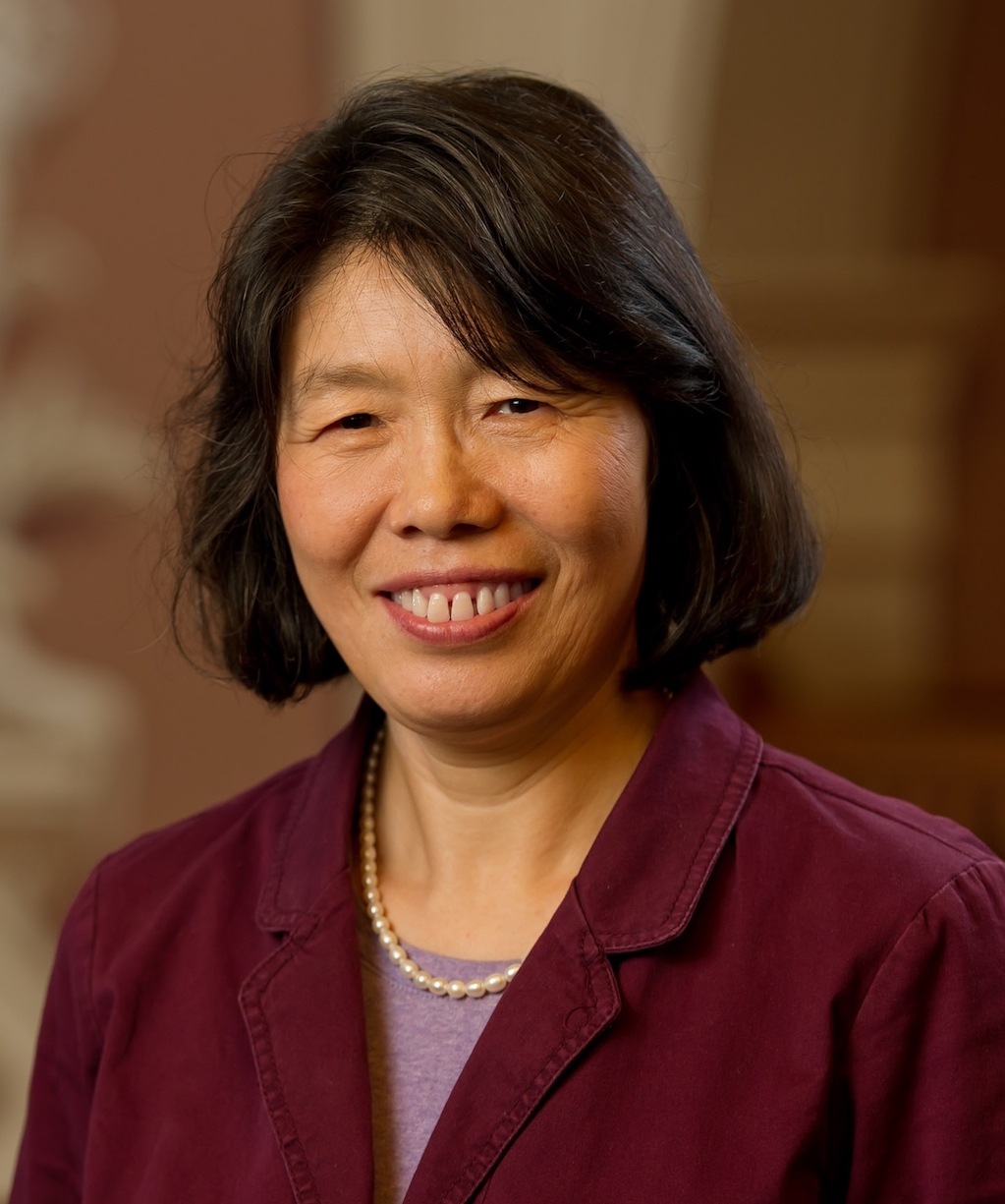}}]{Xiaobo Sharon Hu}
	(S'85-M'89-SM'02-F'16) received her B.S. degree from Tianjin University, China, M.S. from Polytechnic Institute of New York, and Ph.D. from Purdue University, West Lafayette, Indiana. She is Professor in the department of Computer Science and Engineering at University of Notre Dame. Her research interests include real-time embedded systems, low-power system design, and computing with emerging technologies. She has published more than 250 papers in the related areas. She served as Associate Editor for IEEE Transactions on VLSI, ACM Transactions on Design Automation of Electronic Systems, and ACM Transactions on Embedded Computing.  She is the Program Chair of 2016 Design Automation Conference (DAC) and the TPC co-chair of 2014 and 2015 DAC. She received the NSF CAREER Award in 1997, and the Best Paper Award from Design Automation Conference, 2001 and IEEE Symposium on Nanoscale Architectures, 2009.
\end{IEEEbiography}

\end{document}